%% file: main.tex
\documentclass[10pt,a4paper]{article}

\usepackage{natbib}
\setcitestyle{authoryear}%

\usepackage{caption}
\usepackage{subcaption}
\usepackage{graphicx}
\graphicspath{{figures/}}

\usepackage{amssymb}
\usepackage{amsmath} %

\usepackage{algorithmic}
\usepackage{array}
\usepackage{lipsum}
\usepackage{hyperref}

\usepackage{amsfonts}
\usepackage{mathtools}
\usepackage{csquotes}
\usepackage{enumitem}
\usepackage{bbm}

\usepackage{tabularx}
\usepackage{booktabs}
\newcommand{\ra}[1]{\renewcommand{\arraystretch}{#1}}\usepackage{tabularx}
\usepackage{multirow}
\usepackage{multicol}
\usepackage{rotating}

\usepackage{cancel}
\usepackage{tikz}
\usepackage{amssymb}
\usepackage{bbm}
\usepackage{comment}
\usepackage{indentfirst}

\usepackage{hyperref}
\hypersetup{
    colorlinks=true,
    linkcolor=blue,
    filecolor=magenta,      
    urlcolor=cyan,
    citecolor=red
}

\usepackage{listings}

\usepackage{xcolor}

\definecolor{codegreen}{rgb}{0,0.6,0}
\definecolor{codegray}{rgb}{0.5,0.5,0.5}
\definecolor{codepurple}{rgb}{0.58,0,0.82}
\definecolor{backcolour}{rgb}{0.95,0.95,0.92}
\definecolor{LightGray}{gray}{0.9}

\lstdefinestyle{mystyle}{
    backgroundcolor=\color{backcolour},   
    commentstyle=\color{codegreen},
    keywordstyle=\color{magenta},
    numberstyle=\tiny\color{codegray},
    stringstyle=\color{codepurple},
    basicstyle=\ttfamily\footnotesize,
    breakatwhitespace=false,         
    breaklines=true,                 
    captionpos=b,                    
    keepspaces=true,                 
    numbers=left,                    
    numbersep=5pt,                  
    showspaces=false,                
    showstringspaces=false,
    showtabs=false,                  
    tabsize=2
}

\lstdefinestyle{BashInputStyle}{
  language=bash,
  basicstyle=\small\sffamily,
  numbers=left,
  numberstyle=\tiny,
  numbersep=3pt,
  frame=tb,
  columns=fullflexible,
  backgroundcolor=\color{yellow!20},
  linewidth=0.9\linewidth,
  xleftmargin=0.1\linewidth
}

\usepackage{amsthm}

\usepackage{geometry}
\usepackage{setspace}
\begin{document}

\title{Predicting Realized Variance Out of Sample: \\ Can Anything Beat The Benchmark?} %
\author{Austin Pollok \thanks{This paper forms a chapter of my Ph.D. dissertation at the University of Southern California. I am deeply grateful to Christopher S. Jones for his invaluable advice, mentorship, and support throughout the development of this research.} \\
University of Southern California
} %
\date{June 15, 2022} %
\maketitle %

\begin{abstract}
\normalsize
The discrepancy between realized volatility and the market's view of volatility has been known to predict individual equity options at the monthly horizon. It is not clear how this predictability depends on a forecast's ability to predict firm-level volatility. We consider this phenomenon at the daily frequency using high-dimensional machine learning models, as well as low-dimensional factor models. We find that marginal improvements to standard forecast error measurements can lead to economically significant gains in portfolio performance. This makes a case for re-imagining the way we train models that are used to construct portfolios.

\end{abstract}

\setcounter{tocdepth}{2}
\tableofcontents

\clearpage

\section{Introduction}
\label{volatility_chapter}

Volatility as a vague concept refers to both the frequency and magnitude of the ups and downs of a some observable phenomenon over time. Despite the phenomenon being observable, its variation over time is, in fact, not directly observable. Many efforts have been made to estimate, model, and forecast this latent process due to the wide-ranging applications of volatility. Particularly in finance, the range of uses for predicting volatility include risk management, choosing portfolio weights for optimal portfolio construction, pricing and hedging of derivative contracts, and trading on mispricings found in the options market and, more generally, the volatility market.

Historically, one of the most common variables financial economists have focused on studying is the volatility of financial assets' return series, which is an inherently unobserved quantity. We do not sample volatility, we sample prices, which can be tricky. We then aggregate those sampled prices into a measure which summarizes past price fluctuations, ex-post, and possibly predicts future price fluctuations, ex-ante. To further complicate the study of volatility, the frequency and magnitude of ups and downs of prices appears to change throughout time.

Those with a need to understand volatility of returns are faced with answering the following questions:
\begin{enumerate}[label=(\roman*)]
    \item How to \emph{measure} volatility of returns, ex-post, over a given time period? That is, volatility estimation.
    \item How to \emph{predict} volatility of returns, ex-ante, over a given time period? That is, volatility forecasting.
\end{enumerate}

Consider a random variable $R$ modeling the log-gross return\footnote{The log-gross return is also known as the continuously compounded return.} of some financial asset, which we'll often reference as simply the \emph{return}. We can define the volatility as the standard deviation of this return. As a first approximation, we might consider estimating this quantity with sample moments. Given we have observations $\{R_t\}_{t=1}^T$, we define the \emph{sample volatility} as
\begin{equation}
\label{volatility_sample_moments}
\displaystyle
    \hat{\sigma} = \sqrt{\frac{1}{T}\sum_{t=1}^T(R_t - \hat{\mu})^2},
\end{equation}
where $\displaystyle \hat{\mu} = \frac{1}{T}\sum_{t=1}^T R_t$. At short horizons, say daily, the expected return is often small implying its square is effectively zero. Thus it is fair to replace \eqref{volatility_sample_moments} with 
\begin{equation}
    \hat{\sigma} = \sqrt{\frac{1}{T}\sum_{t=1}^T R_t^2}.
\end{equation}
This could easily be improved by using a moving average to discard stale data and take advantage of the observed phenomenon of volatility clustering. Of course, this method is subject to balancing the trade-off between ``staleness'' and noise of the estimator. This trade-off is commonly referred to as the bias-variance trade-off where ``staleness'' refers to bias and variance refers to noise. If you estimate with a shorter sample, your estimator will have more noise despite being more in-tune with the current volatility environment; if you estimate with a longer sample, your estimator will have less noise but will also be less reflective of the current volatility environment. The traditional approach to optimally balance these trade-offs has been to use the (G)ARCH-type models as first introduced by \cite{Engle_ARCH_1982} and \cite{Bollerslev_GARCH_1986}. Additionally, in a world of stochastic observations, particularly stochastic volatility, where the distribution of volatility is changing over time, taking an average over many days as in \eqref{volatility_sample_moments} doesn't make much sense due to the lack of independent and identical distributions of squared returns.

Another simpler, yet effective, approach for the estimation of volatility, ex-post, has been to use range-based estimators, \cite{parkinson_range_based_vol_estimation}, \cite{andersen_bollerslev_the_distribution_of_realized_stock_return_volatility}. In any given trading period, the most commonly recorded prices are the open, close, high, and low price. If we're considering daily trading periods, then the daily range is the difference between the high and low price
\begin{equation}
\label{volatility_range}
    P_t^H - P_t^L;
\end{equation}
and if we want to reduce effects related to price levels, we would need to normalize \eqref{volatility_range}, typically by the closing price
\begin{equation}
    \hat{\sigma} = \frac{P_t^H - P_t^L}{P_t^C}.
\end{equation}
Different flavors and improvements of range-based estimators for volatility have been created and studied by \cite{parkinson_range_based_vol_estimation},  \cite{garman_klass_range_based_vol_estimation},  \cite{rogers_satchell_range_based_vol_estimate}, and  \cite{yang_zhang_range_based_vol_estimate}. A more modern approach to estimating volatility, ex-post, is to use \emph{better} data, where \emph{better} often refers to higher frequency data, which gives rise to the notion of \emph{realized volatility}. We are concerned with answering the following:
\begin{enumerate}[label=(\roman*)]
    \item Are there alternative, high or low-dimensional machine learning, techniques for forecasting volatility that beat the benchmark models out-of-sample?
    \item Can evaluating forecasts from the perspective of an end-user help identify the quality of forecasts?
\end{enumerate}

In this paper we forecast firm-level realized variance out-of-sample, for all firms in the S\&P 500 index, at the daily horizon from 1993 to 2019. To the best of our knowledge, this is largest out-of-sample variance forecasting experiment conducted which also employs machine learning techniques to firm-level variance forecasting. We find, viewed through traditional forecast rankings, it is hard to significantly and unambiguously beat the benchmark models. However, as noted in \cite{patton_vol_loss_functions_2011}, the rankings of volatility forecasts are often ambiguous and can depend on the method of forecast evaluation. This leads us to evaluating the forecasts from the perspective of users of volatility forecasts in the context of forming option portfolios. We find alternative volatility forecasts can lead to economically significant increases to returns, both absolute and risk-adjusted, which is an observed phenomenon in forecasts of the first moment of returns as documented in \cite{campbell_predicting_oos_can_anything_beat_naive_benchmarks}.

\section{Realized Variance}

Due to the unobservable nature of variance, it is common practice to rely on an explicit model of the (G)ARCH-type, for measurement. However, using higher frequency data, there exists a model-free methodology to measure variance which, in turn, provides a natural benchmark for forecast evaluation purposes. Given this model-free methodology for measuring realized variance, one can begin to treat volatility as an observed process and hope to use more traditional forecasting techniques, which treat the variables as observables.

\subsection{Measuring Realized Variance}
Intuitively, a world with availability of tick-by-tick transaction prices and quotes should yield opportunities to estimate volatility with a high degree of precision. If we assume prices evolve in continuous time, then this idea can be made precise \cite{HighFreqFinEconometrics}. Consider working on a daily time horizon with access to a rich data set including minute-by-minute price data, that is, we are no longer restricted to working solely with open, close, high, and low daily price data \footnote{Such a dataset can be constructed from the NYSE TAQ (Trades and Quotes) database.}. A technique, which has become the de-facto standard, to exploit this high frequency data is to use a quadratic variation estimator for an ex-post volatility measurement, as was initially studied in \cite{early_paper_on_realized_vol_andersen_bollerslev}. Such estimators are called \emph{realized volatility} or, for the more mathematically inclined, \emph{realized variation} estimators. 

For a given price process $P$, define \emph{returns} over a given window of time $[t_j, t_{j+1}]$ as the continuously compounded return
\begin{equation}
    \displaystyle
    R_{t_j, t_{j+1}} := \ln\left(\frac{P_{t_{j+1}}}{P_{t_j}}\right).
\end{equation}
Using returns sampled at the $\Delta$ intraday frequency on day $t$, we construct the realized variance estimator by summing the squared returns computed every $\Delta$-period of time
\begin{equation}
\displaystyle
    \label{realized_variance}
    RV_t^{\Delta} := \sum_{j=1}^{\lfloor t/\Delta \rfloor} R^2_{t-1 + (j-1)\Delta, \ t-1 +j\Delta}
\end{equation}
where $1/\Delta$ could be, as it is for our application, 78 for 5-min returns over a 6.5 hour trading day, or 144 for 10-min returns in 24 hour markets.

In the theoretical world of continuous time price evolution, sending $\Delta \rightarrow 0$, which corresponds to continuous sampling, will lead the realized variance estimator to approach the true integrated variance of the price process
\begin{equation}
    \label{realized_variance_approx_integrated_variation}
    RV_t^{\Delta} \underset{\Delta \rightarrow 0}{\longrightarrow} IV_t, 
\end{equation}
in some appropriate sense, where the integrated variance\footnote{Here, we use IV to mean integrated variance, as opposed to implied volatility.} is defined as the quadratic variation of returns over the $[t-1, t]$ period
\begin{equation}
    \label{integrated_variance_is_quadratic_variation}
    IV_t = [R, R]_t - [R, R]_{t-1} \ .
\end{equation}

Under the rather weak assumption of returns evolving as a continuous time semimartingale, we have a method, with theoretical justification, for constructing daily estimates of realized return variation, to any desired degree of precision, using directly observable data. Quadratic variation of a continuous time semimartingale is best interpreted as the actual return variation that transpired over a given period of time, and as such it is the clear target for realized variance measurement. Given that the quadratic variation of a process is the realization of a random quantity, that is difficult to forecast, it serves as a reference for which forecasts should be compared against. Using realized variance as an ex-post variance proxy is completely model-free, requiring nothing more than intraday prices. This allows one to harness the information inherent in high-frequency returns for assessment of lower frequency return variation.

Some issues with high frequency data include the lack of being able to actually achieve the limit. If you take finer and finer samples, you will run into missing data and liquidity effects that will unfavorably bias the estimates. We do not see prices over night which could lead to huge swings in prices between the open and the close and hence jumps in volatility. Further microstructure issues include discreteness of the price grid, as well as bid-ask spreads. This implies high-frequency realized volatility estimators are still noisy, due to a non-negligible error term in the approximation \eqref{realized_variance_approx_integrated_variation}, and likely biased particularly if the sampling frequency is too high. 

\subsection{Forecasting Realized Variance}
Given realized variance is a measurable proxy for the latent quadratic variation, and the associated measurement errors over time are uncorrelated, one could imagine using standard time series models to capture the temporal features of return variance. The most common class of time series models is the autoregressive fractionally integrated moving averaging (ARFIMA(p,d,q)). The class of ARFIMA models does a good job of capturing the statistical properties of logarithmic realized variance \cite{volatility_correlation_forecasting_summary}. Such statistical properties include a long memory dependence structure in variance, as well as logarithmic realized variance being much closer to homoskedastic and approximately unconditionally Gaussian. Such a model, however, leads to a number of practical modeling issues. These issues include the choice of the sampling frequency at which realized variance measures are constructed, the difficulty in disentangling the jumps and diffusive variance components of the realized variance process, and finally, the approach used to best accommodate the indications of \emph{long memory}. If, in good fortune, one can adequately respond to these practical modeling issues, then forecasting is straightforward once the realized variance has been cast in a traditional time series modeling framework, and the model parameters have been estimated. 

Additionally, one must pay attention to how they compute realized variance for a calendar period when the trading day has an official closing. For example, the over-the-counter foreign exchange market has a 24-hour trading period, so this is a minor problem, but this is typically not the case for equity or fixed-income markets. If one uses intraday returns to estimate realized variance, they must be aware the actual object being estimated is the \emph{intraday} realized variance and not the full day's variance. This estimator is not incorporating overnight variance, despite the price process still moving overnight. It is common to see substantial price changes between a market's close and subsequent open, in which case the daily realized variance estimator is missing out on the overnight variance. 

Finally, the preferred sampling frequency can become quite a challenge when the underlying asset is relatively illiquid. If updated price observations are only available intermittently throughout a trading day, then the effective sampling frequency required to generate the \emph{high frequency} returns is lower than the frequency intended to compute the realized variance. To further enhance the problem of an illiquid price series, their bid-ask spreads are typically larger and are more sensitive to random fluctuations in order flow. This implies the associated return series, constructed from the noisy bid-ask spreads, will contain a relatively large amount of noise as well. A simple solution to help reduce the problems of large bid-ask spreads and intermittent trading would be to use a lower sampling frequency, but this too comes at a cost of increasing the measurement error in realized variance. Many of these issues have been previously studied and an even more comprehensive coverage of the literature is presented in \cite{volatility_correlation_forecasting_summary}.

\subsection{Forecasting Firm Volatility}

Financial return series presents stylized facts that induce challenges to classical econometric modeling techniques. These include autocorrelations of the \emph{squared} returns that exhibit strong persistence which can last for long periods of time, as well as return distributions being heavy tailed and highly peaked while displaying a slow convergence to the normal distribution as the horizon increases. Additional stylized facts are summarized in \cite{Cont_Stylized_Facts}. In principle, one hopes volatility models capture the most important stylized facts of stock return volatility; this includes time series clustering, negative correlation with returns known as the leverage effect, log-normality, and long-memory. 
As quoted in \cite{mandelbrot_speculative_prices}, ``large changes tend to be followed by large changes, of either sign, and small changes tend to be followed by small changes,'' which is an informal way of describing time series clustering.

\begin{figure}
\centering
\includegraphics[scale=0.35]{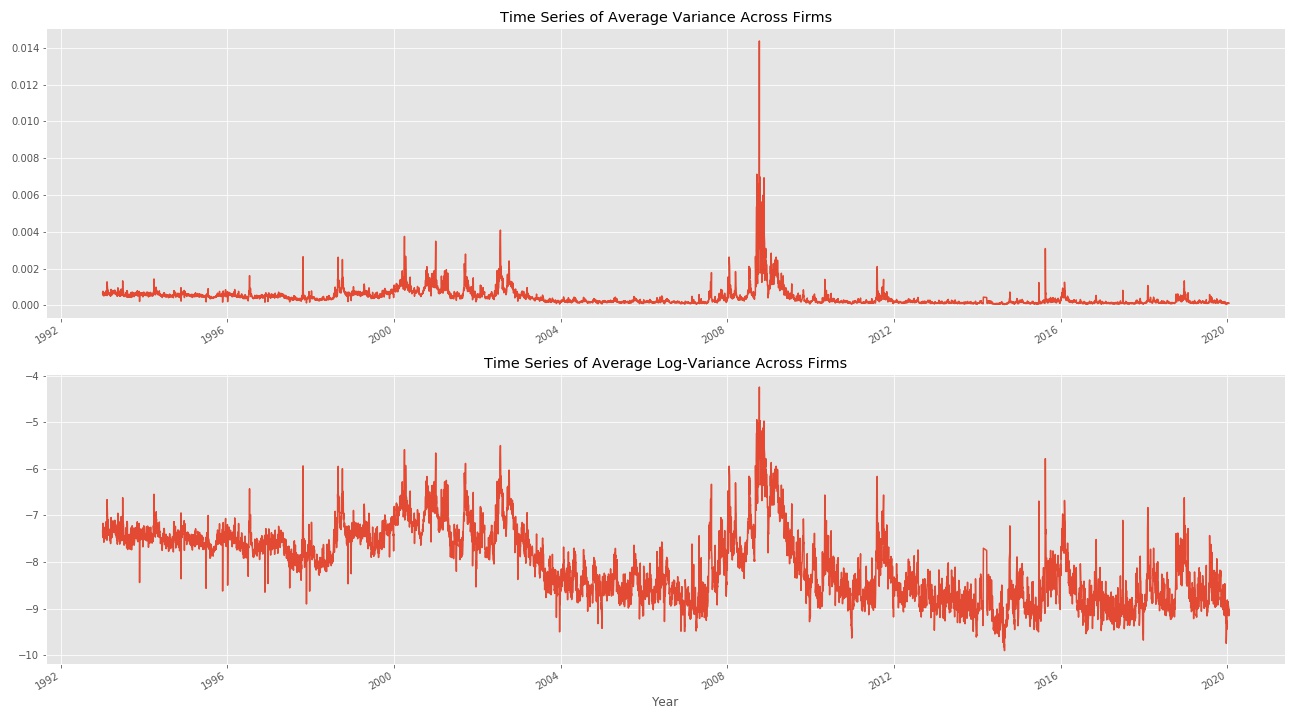}
\caption[time series of average firm realized variance.]{We plot the time series of firm-level realized variance, we note that there appears to be lots of common movement (co-movement) amongst the firms. This suggests the existence of a common factor structure that explains the common movement.}
\label{fig: time_series_avg_rv}
\end{figure}

We see time variation in the firm-level average of realized variances, leading us to believe there exists a common factor structure in variances of individual firms. The time-varying average firm-level volatility points to predictability, in the sense of it not following a random walk. Despite this observation, improving the forecasting ability of the level of volatility of some financial asset's return series turns out to be a highly non-trivial pursuit. As described in \cite{Jones_Brandt}, forecasting the future level of volatility is difficult for multiple reasons. Volatility forecasts are sensitive to the specification of the model; in the case of linear regression this is simply the predictor variable specification. It is important to strike the right balance between capturing the signal in the data and overfitting the noise. Depending on the particular model specification, the volatility forecast can be subject to overfitting the data, rather than extracting the salient features of the data. Given the task of correct model specification, the problem of volatility forecasting is further enhanced by the need to correctly estimate the model's parameters, which can be difficult due to the latent nature of volatility. The further the estimated parameters are from the true parameters, the worse the volatility forecasts will be. Most volatility forecasts depend on current and past levels of volatility which are proxied, or estimated, by noisy variables, which leads, even in the presence of perfectly specified and estimated models, to inherited or possibly amplified uncertainty in the forecast. In our set of experiments, we model the conditional realized variance of returns, and we focus on different model classes, model specification, and out-of-sample forecasting. 
\subsection{Option Returns}
\label{option_returns_chapter}

The breadth of volatility research and innovation in the context of mathematical and econometric volatility modeling, as well as the numerous empirical studies, have lead to several stylized facts in volatility, including time-variation and high-persistence\footnote{Also known as stochastic volatility and volatility clustering.} in asset returns, which in turn has lead to the creation of volatility as an asset class. There are currently many vehicles available for taking implicit or explicit positions in volatility. Traditionally, volatility-based products had a major use case for hedging volatility exposure in portfolios ($\mathbb{Q}$ use cases), however, important new uses show up in asset allocation where portfolio managers take active positions in volatility to capture mispricing or risk premiums ($\mathbb{P}$ use cases). Treating volatility as an asset class can be done because it is easily tradable through various portfolios of exchange-listed options, \cite{evidence_delta_hedged_options_are_bet_on_volatility_bakshi} and \cite{goyal_saretto}, VIX futures and volatility based exchange-traded products (ETPs), or by using other more specialized derivatives \cite{Carr_Lee_Vol_Derivatives} including over-the-counter variance or volatility swaps \cite{Carr_Wu_VRP}, and even correlation swaps \footnote{For a review of equity volatility markets, see \cite{van_tassel_LOOP_in_equity_vol_markets}.}. 

Herein, we take a view on option markets as markets where we can express our view on volatility, where our view is crafted through our forecasts in chapter \ref{volatility_chapter}. These markets can express more than a view on volatility, but also a view on quantities related to volatility such as volatility of volatility, or co-volatility as shown in \cite{Carr_vol_skew_smile_trading}\footnote{Many studies implicitly ask how to trade the implied volatility surface, how to trade statistics associated with realized volatility, and more generally how to trade discrepancies between risk-neutral and statistical distributions, $(f^{\mathbb{Q}}, f^{\mathbb{P}})$, as in \cite{ait_sahalia_discrepancy_of_P_and_Q_densities}.}. As a particular example, one can directly view delta-hedged options returns as being proportional to a volatility risk premium, meaning the returns of someone who trades delta-hedged option positions are rewarded as compensation for bearing volatility risk, as in \cite{evidence_delta_hedged_options_are_bet_on_volatility_bakshi}.

There are many terms to describe the ability to make a profitable trading strategy and they are all associated with different explanations. In particular some common terms that suggest the ability to create a profitable trading strategy are \emph{edge}, \emph{excess expected return}, and \emph{non-zero risk premium}, where a premium refers to having a higher return stream relative to some other return stream. Both edge and excess return imply there is an expectation for profit but don't necessarily provide an explanation as to why such profitability has existed in the past, or why it might hope to persist in the future. This is where the notion of risk premium comes in, it seeks to explain the ability to create a profitable trading strategy by showing that the profits are related to a particular type of risk, which investors dislike. Investors' disdain for this type of risk means they won't hold this risky portfolio unless they are compensated by some required rate-of-return in excess of the risk-free rate, known as the risk premium. This implies the strategy is not an arbitrage, or near-arbitrage, strategy implying the markets are still efficient. Regardless of the terminology, we investigate a hypothetical options trading strategy, and demonstrate, at least as a first approximation, that we can find an edge which is economically significant, and is impacted by our previous methods for volatility forecasting.

Traditionally, financial research on options has been in support of the ``sell-side'' where it is essential to price options and find hedging portfolios that synthesize the option's payoff structure. This follows the vein of \cite{Black_Scholes_pricing_of_options_and_corporate_liabilities} and \cite{Merton_rational_theory_of_option_pricing}, respectively.  Option returns, as concerned from the ``buy-side,'' have, historically, received considerably less attention.

A first theoretical study of option returns, outside the context of pricing and hedging, was in \cite{Option_Markets_Cox_Rubinstein}. In \cite[Section 5.5]{Option_Markets_Cox_Rubinstein} they show how the equilibrium risk and expected return of an option are related to the risk and expected return of the underlying stock in the context of a binomial pricing model. Additionally, they derive relationships between option returns and the CAPM, and, in particular, they show how to calculate the alpha and beta of an option. \cite[Section 5.6]{Option_Markets_Cox_Rubinstein} also do this in continuous time. In \cite[Section 7.8]{Option_Markets_Cox_Rubinstein}, they briefly mention analyzing option risk premium in terms of a factor model. They highlight that market participants who aim to value an option based on the underlying asset or pursue riskless arbitrage profits when market prices differ from the calculated value do not need to study the risks and returns of options. That is, risk-neutral market participants ($\mathbb{Q}-$investors) weren't particularly worried about if an option was a good investment as part of a larger portfolio; such a task was reserved for risk-averse market participants ($\mathbb{P}-$investors). Only later did it become common to include options as part of an investment portfolio, where one needs to consider the risk and return trade-off\footnote{We should mention an additional distinction between returns and prices. Returns represent realized gains or losses on trades, as opposed to prices which are not realized.}. Little empirical research examining these risks, returns, or premiums seem to have been done until \cite{Coval_Shumway_expected_option_returns}.

As noted in \cite[Appendix A]{chernov_understanding_index_option_returns}, prior to the development of markets on index options,there were a few studies on individual security option returns where they assumed the Black-Scholes model was correct and simulated returns from the model. Once the market for index options was developed in the latter-half of the 1980's, many studies, apart from pricing and hedging of index options, tried to better understand the discrepancies between the risk-neutral and statistical probability measures. In particular, \cite{ait_sahalia_discrepancy_of_P_and_Q_densities} used option prices to characterize the risk-neutral density and compare with the underlying's statistical density, and investigate difference between risk-neutral and statistical moments. Additionally, \cite{Jackwerth_recovering_risk_aversion_from_option_prices_and_realized_returns}, tried to better understand the implied volatility smile of S\&P 500 index options which implies out-of-the-money puts are expensive relative to at-the-money puts, and determine if these puts are actually mispriced. Accordingly, many studies analyzed returns to put selling strategies, most at the monthly frequency. 

\cite{Coval_Shumway_expected_option_returns} was the earliest paper studying empirical option risks and returns from an asset pricing point-of-view. They analyzed weekly beta-neutral call and put option returns, as well as straddle returns, and found significantly negative returns, which they claimed is evidence toward additional priced risk factors, rather than a mispricing. Though most of the literature on empirical option returns calculates returns at a monthly horizon,  \cite{jones_nonlinear_factor_analysis_of_index_option_returns} considers daily option returns while using a nonlinear factor model and finds deep out-of-the-money put options have statistically significant alphas and simple put-selling strategies have attractive Sharpe ratios.

More recently, there have been several studies focused on individual equity option returns starting with \cite{goyal_saretto} and \cite{cao_cross_section_equity_returns_idiosyncratic_vol}. They focus on monthly returns from delta-hedged calls and puts, as well as at-the-money straddles, and find these monthly option portfolio returns are predictable due to various measures of volatility such as realized minus implied volatility spreads and idiosyncratic stock volatility, respectively. They postulate their significant option returns come from mispricing or limits to arbitrage.

Most recently, \cite{jones_option_momentum_reversal_seasonality} study the time series and cross-sectional properties of option returns by computing monthly returns on at-the-money straddles on individual
equities. They find option returns exhibit momentum, short-run reversal, and seasonality which further justifies thinking of the option and volatility markets as its own asset class which exhibit a factor structure with similarities to the equities market.

In this study, we focus on returns of at-the-money equity straddles, at the daily frequency, and evaluate the strength of the realized minus implied volatility spread, as in \cite{goyal_saretto}, when we have better forecasts for daily equity volatility.

\subsection{Premiums for Volatility Risk}

From an investor point-of-view, under the $\mathbb{P}-$measure, it is of great importance to understand if volatility risk is good or bad for portfolio returns, and to know how portfolios with exposure to volatility risk should expect to be compensated, or if they should be expected to forfeit returns. One belief is that investors do not like risk, and volatility, as defined by the standard deviation of returns, was one of the first formalized measures of risk. This implies investors do not like volatility which in turn, implies investors \emph{should} be compensated in excess of the risk-free return for bearing such undesired risk. Another belief as to why volatility risk \emph{should} be priced in assets' returns is due to the counter-cyclical dynamics of volatility. It is natural to conjecture that risk-averse investors require compensation for being exposed to volatility shocks and to ask how highly variance risk is priced. 

In response to such needs, a large literature spawned on the analysis of volatility, or variance, risk and how it is priced in a number of different markets. \cite{Carr_Wu_VRP} document variance risk premiums in both the over-the-counter swaps market, as well as the equity market. Similar abnormal returns attributed to variance or volatility risk have been documented in the Treasury market, options market, commodities market, foreign-exchange market, interest-rate swaps market, VIX markets, as well as credit markets as summarized in \cite{ammann_credit_variance_risk_premiums}.

For some time now, traders and academics have observed a discrepancy between the volatility implied by theoretical option pricing models and statistical measures of historical volatility. This discrepancy can be interpreted as risk-averse investors having a different view of volatility than risk-neutral investors. Such a disagreement is one way to define a variance risk premium, and many authors have interpreted the spread between the realized volatility and implied volatility as a measure of volatility risk, and any abnormal returns associated with this measure are capturing a premium associated with volatility risk. There exists many studies that document a negative volatility risk premium. An example of such a study is \cite{Carr_Wu_VRP} who find strong evidence of a negative variance risk premium in expected future stock returns, where they measure variance risk by comparing variance swap rates implied from option prices to realized variances. They also find average at-the-money straddle returns, as well as delta-hedged call and put returns, are large and unconditionally negative, which shows a negative price of volatility risk. It is important to mention \cite{Carr_Wu_VRP} show the variance risk premium is itself a return premium to a class of over-the-counter variance and volatility swaps, which they try to explain with commonly used equity risk factors, such as Fama-French factors; the variance risk premium can also be interpreted as a priced risk factor in equity or option markets\footnote{That is to say, the variance risk premium can be used in the left-hand side and right-hand side of return regressions, which requires it to be explained and to be used for explaining, respectively.} 

Different studies and authors can define variance or volatility risk in different ways, and they can use their defined measure to explain priced risks in various markets, or they can use their defined measure as a return that needs explaining in terms of other risk factors. For the purpose of this study, we consider the volatility risk premium to be the spread between a forecast of future realized volatility and option implied volatility, which provides another forward-looking volatility measure. Implied volatilities are based on the market's forecast of future realized volatilities, extracted from the prices of options on the asset whose volatility we're trying to forecast. One problem with using implied volatility as a forward looking volatility measure is that option prices typically reflect a volatility risk premium due to the fact that options aren't redundant securities, because of limits to arbitrage affecting perfect hedging of volatility risk. This leads to some volatility risk premium being reflected in the computed implied volatilities. It has been documented that the relation between the realized minus implied volatility spread 
\begin{equation}
\left(\widehat{RV}_{t+1} - IV_t, \ \mathbb{E}_t \left[R_{t+1}^{\text{stocks}} \right] \right)
\end{equation}
and future expected stock returns is negative. Additionally, the discrepancy in the realized and implied volatility forecasts has been documented in the cross-section of equity option returns by \cite{goyal_saretto}. They show that the realized minus implied volatility spread 
\begin{equation}
\left(\widehat{RV}_{t+1} - IV_t, \ \mathbb{E}_t\left[R_{t+1}^{\text{options}}\right]\right).
\end{equation}
has a strong positive cross-sectional relation with future expected returns of at-the-money straddles, as well as delta-hedged calls and puts. \cite{goyal_saretto} attribute this phenomenon above to mispricing in the options market. They claim high implied volatility compared to realized volatility are indications of overpriced options, similarly low implied volatility compared to realized volatility indicate underpriced options. 

For a more comprehensive summary of volatility risk in the equity markets see \cite[Chapter 34]{ferson_empirical_asset_pricing} and volatility risk in options and stock markets, see \cite[Chapter 17]{bali_murray_engle_empirical_asset_pricing}. For papers on equilibrium models explaining variance and volatility risk premiums is the S\&P 500 index, see works by  \cite{bollerslev_risk_return_long_run_relations_fractional_cointegration_return_predictability}, \cite{bollerslev_vol_in_equilibrium_asymmetries_and_dynamic_dependencies}, and \cite{bollerslev_expected_stock_returns_and_variance_risk_premia}.

\section{Methodology}
\label{volatility_methodology}

Contrary to most conditional volatility forecasting studies, which test their models on a relatively small number of firms or a small number of distinct asset classes, we investigate our volatility forecasts on the entire cross section of stocks in the  S\&P 500 index at the daily frequency starting in 1/1993 and ending in 6/2019. 

To achieve the most reliable estimate of generalization error while using a static data set, we use a walk-forward out-of-sample forecasting methodology, which to practitioners is known as a back-test, and in the machine learning community as leave-one-out cross-validation (LOOCV).

Given our out-of-sample forecasts, we measure the deviation from the realized variance measurements according to loss functions commonly used in the volatility forecasting literature and theoretically studied in \cite{patton_vol_loss_functions_2011}. We consider mean-squared error, mean-absolute deviation, QLIKE, and Mincer-Zarnowitz regressions. Additionally, we consider, empirically, the differences between aggregating forecast errors across a time series, a cross-section, and the pooled panel of forecasts.

\subsection{Out-of-Sample Analysis}
\label{oos_analysis}

Though often hailed as a virtue in machine learning, ``data mining'' is typically viewed as a demerit in empirical asset pricing, and empirical finance more broadly. Data mining refers to running many repeated experiments to find the one specification that best fits the data, but does not generalize well on unseen data, that is, doesn't fit well on data that was used to estimate the model. One way to account for data mining is to adjust the statistical significance threshold according to the number of parameter searches undertaken, this is known as the problem of multiple-testing in machine learning \cite[Section 18.7]{elements_of_statistical_learning} and was studied in the finance literature by \cite{harvey_multiple_testing_cross_section_of_expected_returns}. Another reasonable response to the problem of data mining is to use an out-of-sample (OOS) walk-forward validation approach, which financial practitioners call a \emph{back-test}. For more on the issues of data snooping, see \cite[Section 32.1.3.2]{ferson_empirical_asset_pricing}. 

True out-of-sample tests are often not feasible because we're working with a static data set, so we must rely on pseudo-OOS tests, which imitate real-time forecasting on historical data, ex-post. We must concede that pseudo-OOS tests can still be subjected to p-hacking, multiple testing, and data mining as described in \cite{goyal_saretto_p_hacking}. We perform a rolling out-of-sample analysis on our panel of data \eqref{empirical_data_setup_out_of_sample}. The vague idea of this error estimation technique is to forecast exactly as we would use the forecast in real-time on live, unseen data. More specifically, on every day, and for every firm, we split our sample into a set of data that is used to estimate our forecasting model and a set of data that is used to evaluate the forecast against the measured response variable. Next we roll the sets forward by one period and repeat. Such sets have varying names depending on the academic circle. The set used to estimate the model can be called the estimation set, training set, or learning set. The set used to evaluate the the model's forecast against the measured variable can be called the out-of-sample set, test set, or validation set.

An additional reason for using a walk-forward out-of-sample analysis is to account for structural change in the data. Structural changes can come from two sources, either the ambient environment changing or environment participants learning from past data and adapting their behavior going forward. A particularly simple technique for addressing the problem of structural change is to use a rolling window estimation set where data prior to a certain time period is completely discarded, or used with a weight of zero. Another data weighting scheme is to use exponential weighting that gradually down-weights old data. It is not exactly clear which schemes make the most sense in financial applications. With these structural change concerns in mind, we construct our estimation set by employing a rolling window approach which gives a simple technique for estimating time varying conditional volatilities for a given estimation window $[t-W, t]$. The window length, $W$, directly determines the bias-variance trade-off of the forecast, with larger values of $W$ increasing the forecast bias, due to nonstationarity of financial return data, while reducing the forecast's variance.

More specifically, we estimate our model for conditional variance on every day $t$, and for every stock $i$ in the sample\footnote{There are 1078 unique firms in our sample starting from 1993 and ending in 2019.}, using $W = 250$ days of observations of the predictor variables. Every time $t$ we estimate our forecast using data from $t-W$ to $t-1$, producing a data set of the form
\begin{equation}
\label{empirical_data_setup_out_of_sample}
    \bigg\{\underbrace{(\mathbf{x}_{t-W}, y_{i, t-W+1}), \dots,(\mathbf{x}_{t-2}, y_{i, t-1}),}_{\text{estimation (training) set}} (\mathbf{x}_{t-1}, y_{i, t}), \underbrace{(\mathbf{x}_{t}, y_{i,t+1})}_{\text{OOS (test) set}}\bigg\}\footnote{We do not have a ``standard'' machine learning setup for every stock and for every day. The observations $(\mathbf{x}, y)$ are not independent and identically distributed, but rather have dependence and a time-varying distribution between the observations.},
\end{equation}
which gives a forecast $\hat{y}_{i,t+1} = f(\mathbf{x}_t; \hat{\beta}_t)$ for the next period $t+1$, which we will compare against the next period's observed value $y_{i,t+1}$.

To preserve an out-of-sample analysis, it is important to recognize the information we have available at time $t$ when we estimate our forecast. Another way of saying this is we must preserve adaptedness of our forecasts to the available filtration $\mathbb{F} = \{\mathcal{F}_t\}$. We're trying to forecast the $t+1$ daily realized variance at time $t$ using information up to $3:55$pm ET. We use data before $3:55$pm ET in order to allow for a tradeable strategy employing the variance forecasts. Using data before the close of the trading day, $4$pm ET, allows us to place trades on day $t$ using our estimated forecast for $t+1$.

The order of operations is as follows:
\begin{enumerate}
    \item Estimate a forecast model which is trained between the close in period $t-1$ and before $3:55$pm ET in period $t$.
    \item Measure the predictor variables at time $3:55$pm ET, $\mathbf{x}_{t} \in \mathcal{F}_t^{3:55}$, and plug them into our estimated model with produces a forecast $\hat{y}_{i,t+1} = f(\mathbf{x}_t; \hat{\beta}_t) \in \mathcal{F}_t^{3:55}$.
    \item Place trades before the close of the trading day at $4$pm ET in period $t$ to be realized in the next period $t+1$.
    \item Measure the quality of the forecast in period $t+1$ based on $\hat{y}_{i,t+1}$ and the measured $y_{i,t+1}$.
\end{enumerate}

Because we estimate our model between the close of period $t-1$ and before $3:55$pm ET in period $t$, we cannot include the observation $(\mathbf{x}_{t-1}, y_{i,t})$ in our estimation set. We do not have access to $y_{i,t}$, which we wouldn't have access to until after the close of period $t$, even though we do have access to $\mathbf{x}_{i, t-1}$. Including  $(\mathbf{x}_{t-1}, y_{i,t})$ in our estimation set would destroy the out-of-sample analysis by introducing look-ahead bias to a trading strategy making use of our forecast.

Another subtle issue with our sample is not all stocks remain in the sample the entire estimation or out-of-sample set. As firms can enter and exit the S\&P 500 index based off market valuations, the panel of data becomes unbalanced. Because we are working with an unbalanced panel at each point in the sample \eqref{empirical_data_setup_out_of_sample}, we choose to apply a filter which drops any firms that enter or exit the estimation set. This allows us to avoid having to impute values for firms that enter or exit the estimation set, and does not use introduce any look-ahead bias.

\subsection{Out-of-Sample Performance Measurement}
\label{vol_oos_performance_measurement}

Using the walk-forward out-of-sample methodology to forecast the daily realized variance yields a sequence
\begin{equation}
\label{rv_measured_forecasted}
    \bigg\{(y_{i, t+1}, \hat{y}_{i, t+1}); 1 \leq t \leq T, 1\leq i \leq N\bigg\}
\end{equation}
of daily measured realized variance and the associated forecast for that day, for every firm. Let $N$ be the number of firms and $T$ be the number of trading days in our sample\footnote{In our sample, $T=6350$, $N=1078$.}. There exists familiar techniques for measuring volatility forecast error, which often vary depending on the empirical properties of the data, though most of these techniques do not address how to aggregate forecast error measurements in a large panel of forecasts, as we have. 

In the volatility forecasting literature, numerous authors have expressed concern that a few extreme observations may have an excessively large impact on the outcomes of forecast evaluation and comparison tests. The common response to such concerns is to use loss functions that are less sensitive to large observations, such as mean absolute deviation or proportional error, instead of the traditional mean squared error loss function. \cite{patton_vol_loss_functions_2011} theoretically shows this approach can still lead to incorrect inferences and selection of inferior forecasts. 

The forecaster must then consider how different loss functions penalize deviations differently. The mean squared error is more sensitive to outliers relative to other commonly used loss functions. QLIKE\footnote{The QLIKE loss function is defined as $\mathcal{L}(y, \hat{y}):= \log(\hat{y}) + \frac{y}{\hat{y}}$.}, however, is an asymmetric loss function, viewed as a function of forecast value. The far left tail of the QLIKE loss function is far more sensitive, yet less sensitive to the far right tail of the loss function relative to MSE. This property of QLIKE partially solves the problem of volatility loss functions being dominated by a few large observations by being robust to extreme observations in the right tail while sacrificing robustness in the left tail. Being a non-symmetric loss function, QLIKE penalizes positive and negative loss values differently which leaves the ranking of forecasts by QLIKE to incorrectly favor positively biased forecasts. Further problems arise through distortions in the rankings of competing forecasts when using a noisy volatility proxy. Our realized variance volatility proxy still contains noise, despite being less noisy than traditional proxies such as squared returns or ranged-based estimators. \cite{patton_vol_loss_functions_2011} demonstrates the MSE and QLIKE loss functions are, theoretically, robust to noise in the volatility proxy for the latent volatility process. \cite{Meddahi_justification_for_MZ_regression} shows the ranking of forecasts on the basis of $R^2$ from Mincer-Zarnowitz (MZ) regressions\footnote{MZ regressions take the form $y_t = a + b \hat{y}_t + \varepsilon_t$.}, where one regresses a noisy variance proxy on the variance forecast, is robust to noise in the true variance proxy, but still might not handle the few extreme observations problem we see in volatility forecasting. \cite[Section 3.3]{corsi_har_rv_model} also uses MZ-regressions for evaluating forecast performance.
Given the data \eqref{rv_measured_forecasted}, we estimate the out-of-sample prediction error using a squared loss function, absolute loss function, and the QLIKE loss function analyzed in \cite{patton_vol_loss_functions_2011} and applied in \cite{patton_loss_functions_volatility}.  We also use the $R^2$ from MZ regressions \cite{MZ_Regressions} where we regress the ex-post measurement onto the ex-ante forecast to measure performance of the forecasts, as well as the bias. For these MZ-regression based forecast evaluations, unbiasedness of the forecasts requires an intercept of zero, and a slope of one. MZ-regressions rank volatility forecast models by their ability to \emph{explain} subsequent realized volatility measurements, as measured by $R^2$. When using the $R^2$ of MZ-regressions to rank forecasts, recall
\begin{equation}
R^2 = 1 - \frac{Var(\hat{\varepsilon})}{Var(y)} = 1 - \frac{\sum_i \hat{\varepsilon}_i^2}{\sum_i (y_i - \bar{y})^2},
\end{equation}
where $\hat{\varepsilon} = y_i - \hat{y}_i$ which leads to the ratio $\frac{\sum_i \hat{\varepsilon}_i^2}{\sum_i (y_i - \bar{y})^2}$ being interpreted as the fraction of unexplained variance, that is, variance not explained by a model. For $R^2$ to increase, variance of the residuals needs to decrease. In the extreme case of perfect over-fitting where $\hat{y}_i = y_i$, for all $i$, we have $R^2 = 1$. If we consider a constant baseline model $\hat{y}_i = \bar{y}$, then $R^2 = 0$, with models forecasting worse than the mean leading to $R^2 < 0$. 

Regardless of the choice of forecast error measurement, it has been known, despite the obvious increase in validity, that a walk-forward out-of-sample evaluation of forecast errors provides evidence that is hard to interpret. For example, \cite{campbell_predicting_oos_can_anything_beat_naive_benchmarks} find when predicting the first moment of returns, a variable can have economic predictive power, yet still fail to outperform a naive benchmark when predicting using a walk-forward validation test. All of the different methods for forecast evaluation have imperfections, which is why in section \ref{option_return_forecast_results} we rank volatility forecasts by evaluating returns of option portfolios formed on the basis of the forecast.

\subsection{Option Portfolios}

In order to evaluate the economic significance of our volatility forecasts from chapter \ref{volatility_chapter}, we must compute the spread between our realized volatility forecast and the market's implied volatility forecast. We will use this signal to form cross-section portfolios made up of \emph{daily}, at-the-money, delta-neutral straddles. We do not decompose the investable universe based on time-to-maturity of the available straddles. Our formation of these cross-section portfolios uses the \emph{portfolio-sort} methodology, which is a simple nonparametric technique for estimating the future conditional expected rate-of-return, conditional on a \emph{sorting} variable. All variables are appropriately lagged to avoid any look-ahead bias, data snooping, and leakage into our portfolio return estimate. Finally, we evaluate the portfolio performance by looking at its time series and computing the average rate-of-return, both absolute and risk-adjusted.

In empirical studies where a trading strategy is being investigated, the importance of removing all possible leakage of data and look-ahead biases cannot be overly emphasized. One source of look-ahead bias is not ensuring variables are appropriately lagged. Specifically, data needs to be shifted backwards to ensure information used to make a forecast for today could not have been known after yesterday. That is to say all variables used for forecasting must be appropriately adapted to the filtration. When we sort portfolios, it is imperative that the sorting variable is adapted to the time period in the \emph{denominator} of returns. If 
\begin{equation}
R_t = \frac{P_t}{P_{t-1}}-1,
\end{equation}
then the sorting variable 
\begin{equation}
X_{t-1} \in \mathcal{F}_{t-1}
\end{equation}
must be $(t-1)-$measurable because we form portfolios in the previous period and then observe how they perform over the current period. For the purpose of our experiments, we have forecasts for firm-level realized and implied volatilities\footnote{We ensure $IV$ and $\sqrt{RV}$ are in the correct units. Because $IV$ is annualized, we need to de-annualize by dividing by $\sqrt{250}$, to make the units daily rather than yearly, to match the units of realized variance.}
\begin{equation}
\widehat{RV}_{t}, \ IV_{t-1} \in \mathcal{F}_{t-1}^{\text{3:55pm}}
\end{equation}
which gives us enough time to plug our predictor variables into our trained model and get our forecast for tomorrow's realized-variance, calculate a sorting variable, sort options, and subsequently form portfolios whose returns will be realized over the following period. We use a measure of the spread between realized and implied volatility as our portfolio sorting variable,
\begin{equation}
X_{t-1} := \text{VRP}_{t-1} = f\left(\widehat{RV}_t, I_{t-1}\right) \in \mathcal{F}_{t-1}^{\text{3:55pm}},
\end{equation}
where $f(x,y) = x-y, \ \frac{x}{y}$, or $\ln(\frac{x}{y})$.

Upon forming the properly lagged sorting variable, we form portfolios from the assets in the investable universe as a way to test the predictive power of the sorting variable on the conditional expected future return\footnote{In \href{https://cattaneo.princeton.edu/papers/Cattaneo-Crump-Farrell-Schaumburg_2020_RESTAT.pdf}{Characteristic-Sorted Portfolios: Estimation and Inference}, \cite{cattaneo_characteristic_sorted_portfolios_statistical_perspective} investigate the commonly used nonparametric technique of portfolio sorts from a statistical perspective, and show it is a nonparametric estimator for conditional expected returns, as well as derive many of its statistical properties.}. This also yields a hypothetical naive trading strategy.

For each period, $[t_{i-1}, \ t_{i}]$, we consider an investable universe, and weights $w_{i,t-1}$ constructed at the beginning of the period, for simplicity we use equal weights. Next, we select, or sort, assets from the investable universe based on their rank ordering, i.e. binning, according to some characteristic, $X_{t-1}$. It is important to realize this particular classification of assets is based on a rank-ordering from quantiles according to a characteristic variable, and can be further generalized by other machine learning methods used to discriminate based on different characteristics. Given we've sorted the investable universe by rank ordering, we form portfolios by multiplying the returns of the selected groups, typically the highest and lowest quintile or decile, by weights $w_{i,t-1} = \frac{1}{N}$, where $N$ is the number of assets we've selected by binning. This yields a portfolio return for every point in time, which gives us a time series of portfolio returns and allows us to analyze the portfolio's performance.

When the investable universe is options on the universe of S\&P 500 stocks, there are lots of ways to select options, and there are common heuristics researchers use with regards to the strike or moneyness and term-structure of all available options.

\section{Models}
\label{models_section}

Historically, volatility modeling has evolved along two distinct paths corresponding to the statistical, $\mathbb{P}$, and risk-neutral, $\mathbb{Q}$, probability measures. The statistical $\mathbb{P}$ probability measure path has typically followed discrete-time (G)ARCH-type models as in \cite{Engle_ARCH_1982} and \cite{Bollerslev_GARCH_1986}, while the risk-neutral $\mathbb{Q}$ probability measure path has typically followed continuous-time dynamics modeling as in \cite{ait_sahalia_implied_stochastic_volatility_models}. The (G)ARCH-style literature has typically focused on explicitly modeling the stylized facts of volatility, such as autocorrelation and persistence.

Without explicitly focusing on modeling stylized facts of volatility in-sample we, instead, focus on out-of-sample forecasting of realized variance utilizing both high and low-dimensional machine learning models.

\subsection{Heterogeneous Autoregressive Model of Realized Variance}
\label{corsi_har_subsection}

A volatility forecasting model that has been demonstrated to have good forecasting performance, on a small sample of assets, is 
the heterogeneous autoregressive (HAR) model of \cite{corsi_har_rv_model}. This model is a simple alternative to the seminal (G)ARCH-type models of \cite{Engle_ARCH_1982} and \cite{Bollerslev_GARCH_1986}; it uses ordinary least squares to regress a firm's daily realized volatility on its own lagged realized volatility at various horizons. Corsi finds by exploiting the information content in high frequency data to construct variance measures, and by removing the need for numerical
optimization, the HAR model out-performs classical (G)ARCH-type models which explicitly model stylized facts of return volatility. 

Using high-frequency intraday data to construct realized variance estimators, Corsi \cite{corsi_har_rv_model} makes a simple improvement to this realized variance estimator by modeling realized variance using an autoregressive model and lagged realized variance estimates over daily, weekly, and monthly horizons,
\begin{equation}
    \label{corsi_model}
    \displaystyle
    RV_{t+1}^{(d)} = c + \beta^{(d)} RV_t^{(d)} + \beta^{(w)}RV_t^{(w)} + \beta^{(m)}RV_t^{(m)} + \varepsilon_{t+1}, 
\end{equation}
where, for $t$ in days, $RV_t^{(d)}$ is the usual ex-post daily realized variance estimator \eqref{realized_variance}, and $\displaystyle RV_t^{(w)}, \ RV_t^{(m)}$ are equal-weighted averages of daily realized variance over the past week and month, respectively,
\begin{align}
    \label{realized_variance_weekly_avg}
    & RV_t^{(w)} = \frac{1}{5}\left( RV_t^{(d)} + RV_{t-1}^{(d)} + \cdots + RV_{t-4}^{(d)} \right), \\
    \label{realized_variance_monthly_avg}
    & RV_t^{(m)} = \frac{1}{22} \left( RV_t^{(d)} + RV_{t-1}^{(d)} + \cdots + RV_{t-21}^{(d)} \right).
\end{align}
We can loosely view \eqref{corsi_model} as a time series counterpart to a three factor stochastic volatility model where the factors are the past realized variances viewed at different frequencies. 

Equation \eqref{corsi_model} has a simple autoregressive structure in realized volatility with the additional feature of considering volatilities over different horizons, this observations leads Corsi to name this model heterogeneous autoregressive of realized volatility, HAR(3)-RV model, which we simply refer to as HAR. Another nice feature of \eqref{corsi_model} is the simplicity of the implementation method. A standard linear regression, fit with the method of ordinary least squares, will suffice to estimate a volatility forecasting model.

If we assume log-prices move as a continuous time semimartingale, then the theoretical volatility over the course of the day is the quadratic variation of the semimartingale \eqref{integrated_variance_is_quadratic_variation}, coined integrated variance, which is well approximated by \eqref{realized_variance}, as seen by a corresponding limit theorem \eqref{realized_variance_approx_integrated_variation}. Corsi empirically showed we can use an autoregressive model depending on different horizons of realized variance \eqref{corsi_model} to approximate \eqref{realized_variance}, both as an ex-post estimator and an ex-ante forecast. Due to the simplicity and forecasting success of the HAR model, we choose to use \eqref{corsi_model} as our baseline for forecasts of daily realized variances \footnote{We consider other specifications of the HAR model, such as using returns squared as the dependent variable, different frequencies of model re-estimation, different horizons for the predicted realized variances, as well as some nonlinear functions of the Corsi regressors, with little to worsening differences in performance.}.

Viewing \eqref{corsi_model} more generally as a factor model, one can speculate as to whether it would be of further use to include additional factors such as various measures of jumps, overnight or holiday volatility indicators, factors for news announcements, implied volatility estimates from option prices or volatility swap contracts, etc. In addition to adding more regressors, we could also try to incorporate some nonlinear effects by changing the underlying linear model to a nonlinear one. We begin to tackle some of these questions in the following sections.

\subsection{High-Dimensional Regularized Models of Realized Variance}
\label{high_dimensional_regularized_models_of_rv}

We know when measuring volatility all estimators are noisy. The realized variance of a stock's daily return, $RV_{i,t}$, as demonstrated by \cite{corsi_har_rv_model}, is affected by its own previous realized variance $RV_{i, t-1}$ at different horizons. Additionally, as first described by \cite{black_leverage_effect}, volatility is related to returns, $R_{i, t-1}$. Black documents the tendency of variances to go up after the market return goes down; specifically, volatility responds to a negative return much greater than a positive return of the same magnitude. Black postulated this effect was caused from a drop in a stock's price, which changes the capital structure of the firm by increasing the debt-to-equity ratio since a falling price decreases equity value. This explanation was coined as the leverage effect. This has been modeled in continuous time by \cite{Carr_Wu_Leverage_Effect}, and empirically by \cite{Lo_Leverage_Effect}, where they showed the leverage effect is not actually due to firms' leverage. Nevertheless, the leverage effect has come to be known as the asymmetry of magnitudes in upward and downward movements of returns, which is commonly observed in equity markets. We hypothesize firm-level realized variance is also related to the previous realized variance and returns of other related stocks in the cross-section. How these stocks in the cross-section become related could be through a wide variety of mechanisms, or common factors, some more macroeconomic, others more micro-structure related. We hope to capture some of the related stocks whose past volatility and returns have some influence over $RV_{i,t}$, that is we hope to capture \emph{sparse-signals} in the cross section of realized volatility. This is motivated by \cite{chicno_sparse_signals_cross_section} who study the sparse-signals in the cross-section of stock returns. Additionally by adding more variables and information we hope to reduce noise and increase bias in a favorable direction. By exploiting the trade-off between bias and variance we hope to reduce forecast error. With the above observations in mind, we model realized variance using the cross-section of returns and realized variances at different horizons, which forces the forecasting problem into a high-dimensional framework.

In high-dimensional settings, where the number of predictor variables isn't small relative to the number of observations, or when there are more predictors relative to the number of observations, predictions based on ordinary least squares estimates are often unreliable. When the number of covariates is large relative to the number of observations, ordinary least squares overfits the data by tipping the parameters $\hat{\beta}^{OLS}$ to fit noise rather than true signal. In the case where the number of input variables is strictly larger than the number of observations, there are an infinite number of of solutions to $\hat{\beta}^{OLS}$, as ordinary least squares estimates are not unique. In particular, there are an infinite number of solutions that fit the training data perfectly, but much of this fit is from fitting the $\varepsilon_{i,t}$ in the observation $y_{i,t}$ which we're modeling as $y_{i,t} = f(\mathbf{X}_{i,t-1}; \beta_{t-1}) + \varepsilon_{i,t}$. One approach to help reduce the problem of a model fitting the noise in an observation, especially in a high-dimensional setting, is to \emph{regularize} a model. For our application, regularization loosely means constraining the model's parameters in a particular way. Regularization can reduce variance, but at the cost of biasing the forecast. That is, a forecast which fits the training data without much regularization can lead to a forecast with high estimation error and low bias. Increasing the \emph{amount} of regularization, as controlled by a hyperparameter, or the \emph{type} of regularization, as controlled by the regularization function, can lead to a forecast with lower estimation error and a higher bias. 

To model realized variance in this high-dimensional setting, we use the least absolute shrinkage and selection operator (LASSO) of \cite{tibshirani_lasso} to simultaneously select and shrink regression coefficient estimates, which is used to prevent overfitting. The LASSO estimates parameters by minimizing the objective function 
\begin{equation}
\label{lasso_equation}
\mathbf{\hat{\beta}}_{\lambda}^{LASSO} = \operatorname*{argmin}_{\mathbf{\beta}} \left( (\mathbf{y}-\mathbf{X}\mathbf{\beta})^T(\mathbf{y}-\mathbf{X}\mathbf{\beta}) + \lambda ||\mathbf{\beta}||_1\right).
\end{equation}
The LASSO parameter estimates $\mathbf{\hat{\beta}}_{\lambda}^{LASSO}$ are non-linear in the dependent variable, unlike the ordinary least squares parameter estimates, and there is no closed-form solution. To compute $\mathbf{\hat{\beta}}_{\lambda}^{LASSO}$ requires solving a quadratic programming problem, for which there are a number of efficient algorithms to solve \ref{lasso_equation}, as described in \cite[Sections 3.4.4 and 3.8]{elements_of_statistical_learning}. The $L^1$ penalty function in \ref{lasso_equation} induces shrinkage of coefficients towards zero, as well as \emph{sparse} coefficient estimates. That is, only a small number of the coefficients in $\mathbf{\hat{\beta}}_{\lambda}^{LASSO}$ will be nonzero\footnote{We should note the LASSO can run into problems with correlated variables.}.

For every stock in S\&P 500, from 1993 to 2019, we use all cross-sectional stocks' lagged daily, weekly, and monthly realized variance and return estimates\footnote{Computed in the same fashion as \eqref{realized_variance_weekly_avg} and \eqref{realized_variance_monthly_avg}.}, measured at at $3:55$pm ET, as predictor variables to forecast the next day's firm-specific realized variance
\begin{equation} 
\label{sparse_signals_realized_variance_model}
\begin{split}
\displaystyle
\widehat{RV}_{i,t+1} =  f\bigg( & \{RV_{j,t}^{3:55, (d)}\}_{j=1}^{500}, \ \{R_{j,t}^{3:55, (d)}\}_{j=1}^{500}, \\ 
& \{RV_{j,t}^{3:55, (w)}\}_{j=1}^{500}, \ \{R_{j,t}^{3:55, (w)}\}_{j=1}^{500}, \\
& \{RV_{j,t}^{3:55, (m)}\}_{j=1}^{500}, \ \{R_{j,t}^{3:55, (m)}\}_{j=1}^{500};  \quad \mathbf{\hat{\beta}}_t\bigg).
\end{split}
\end{equation} As discussed in section \ref{volatility_methodology}, the training sample is based on $W$-many observations from a window of time, $[t-W, t-1]$. If our forecasting model, $f(\mathbf{x};\mathbf{\beta})$, is a penalized linear model such as LASSO, Ridge, or Elastic-Net regressions, also called \emph{shrinkage} models, we need a methodology for how to best choose the hyperparameters in the context of a time series forecast. In our setting, $\lambda$ in \eqref{lasso_equation} and the window parameter $W$ are considered hyperparameters.

One data-driven methodology to choose hyperparameters is \emph{cross-validation}. In any prediction or forecasting problem, one fits a model to a particular data set, that is, estimates a model's parameters to make the output of the model optimally close to the data set. Classical statistics often emphasizes the \emph{in-sample error} which is a theoretical quantity commonly formulated as $MSE(\hat{y}), RMSE(\hat{y})$, or $MAD(\hat{y})$, which one can estimate empirically. Another theoretical quantity which measures a model's performance on unseen data, is know as the \emph{out-of-sample (generalization) error}. Out-of-sample error can be estimated with the technique of cross-validation
\begin{equation}
\label{OOS_error_estimate}    
    \varepsilon_{OOS} := \mathbb{E}\left[error(\hat{Y}, Y)\right] \approx \hat{\varepsilon}_{OOS}^{CV}.
\end{equation}
There exists many variants to cross-validation that can reduce the variation in the out-of-sample error estimate \eqref{OOS_error_estimate}, including a \emph{resampling} methodology coined $K-$fold cross-validation. The method of cross-validation can also be used to estimate the theoretically optimal set of model hyperparameters. We use the method of cross-validation to simultaneously select the best set of hyperparameters as well as estimate the out-of-sample error; this double cross-validation is referred to as \emph{nested cross-validation}. Specifically, a particular out-of-sample error estimate depends on the choice of model hyperparameters,
\begin{equation}
\label{nested_CV}
    \hat{\varepsilon}_{OOS}^{CV}(\lambda) \approx \mathbb{E}\left[error(\hat{Y}^{\lambda}, Y) \right] ,
\end{equation}
and one wants to choose the set of hyperparameters that that minimizes the out-of-sample error estimate
\begin{equation}
    \hat{\lambda}^{*, CV} = \operatorname*{argmin}_{\lambda} \hat{\varepsilon}_{OOS}^{CV}(\lambda) \approx \lambda^*,
\end{equation}
computed with, say, $K-$fold cross-validation, where 
\begin{equation}
    \lambda^* := \underset{\lambda}{\arg\min} \mathbb{E}\left[error(\hat{Y}^{\lambda}, Y) \right].
\end{equation}

For every day $t$, and for every stock $i$ in the sample\footnote{There are 1078 unique firms in our sample starting from 1993 and ending in 2019.}, using $W = 250$ days of observations of the predictor variables, we estimate our forecast using data from $t-W$ to $t-1$, using a data set of the form
\begin{equation}
\label{empirical_data_setup_out_of_sample_and_hyperparamater_estimation}
\begin{split}
    & \bigg\{\underbrace{(\mathbf{x}_{t-W}, y_{i, t-W+1}), \dots,(\mathbf{x}_{t-W+j}, y_{i, t-W+1+j})}_{\text{estimation (training) set}}, \\
    & \underbrace{(\mathbf{x}_{t-W+j+1}, y_{i, t-W+1+j+1}), \dots, (\mathbf{x}_{t-2}, y_{i, t-1})}_{\text{validation set}}, (\mathbf{x}_{t-1}, y_{i, t}), \\
    & \underbrace{(\mathbf{x}_{t}, y_{i,t+1})}_{\text{OOS (test) set}}\bigg\}
\end{split}
\end{equation}
for some $j \leq t-2$. The out-of-sample set is used for evaluating forecasts for the estimated model, and the validation set is for evaluating forecasts for different model configurations, that is, different hyperparameters; finally the estimation set is used to estimate the model parameters for a given hyperparameter. For more on nested cross-validation, see \cite[Section 7.10]{elements_of_statistical_learning}.

It is worth noting that in regularized models, there is an issue of how to track structural changes in the parameters of the forecast model, and also how to estimate and adapt the hyperparameters over time. There isn't a good reason to fix the hyperparameters over time which necessitates a methodology for a data-driven method to estimate how the hyperparameters changes over time. Simply re-estimating the penalty hyperparameters every period in a rolling windows walk-forward out-of-sample validation may be computationally too expensive. Concerns of nonstationary data, that is, structural change, raise questions about the suitability of cross-validation methods for both model parameter and hyperparmeter estimation. Implementations of $K-$fold cross-validation assume the temporal position of the validation fold relative to the training folds is irrelevant. Validation folds can be drawn from the data that is older than some of the training data, which isn't a problem with stationary data. When data is nonstationary, it isn't clear that $K-$fold cross-validation is the best methodology for estimating parameters, because the direction of time can matter. So far, there has been little research in tackling this question and is an important open question, which we leave for future work.

We should mention \cite{Carr_ML_RV} use ridge regression, random forests, and feed-forward neural networks to predict realized variance of the S\&P 500 index (SPX). They find that machine learning methods have marginal improvement when predicting the realized variance of SPX using option prices as predictor variables. Instead, they find greater improvements when predicting a risk premium, that is, the difference between the realized variance of SPX and a VIX-styled volatility index, which they synthesize in the paper. 

There are a few other variants of the high-dimensional regularized models of realized variance that we investigate, but nothing seems to deviate much from the LASSO specification\footnote{We don't change the left or right-hand side regression variables, but we do change the functional form, $f$, of the model to ridge regression and the elastic-net. To reduce computational time, we also only retrain our model every twenty trading days, so approximately monthly.}.

Finally, as mentioned in \cite[Section 2.4]{nagel_ml_in_asset_pricing}, regularization can be interpreted from a Bayesian perspective. An open research agenda, as laid out in \cite{nagel_ml_in_asset_pricing}, is to look for links between regularization and economic restrictions, which requires the Bayesian interpretation of regularization.

\subsection{Low-Dimensional Statistical Factor Models of Realized Variance}
\label{Low-Dimensional Statistical Factor Models of Realized Variance}

In general, consider observing the time series of a measurable quantity on a collection of entities, such as test assets.If the test assets' time series tend to jointly move together, then this suggests there may be a common factor structure that is driving this common movement or \emph{synchronization} of the time series. By a common factor structure, we mean a set of variables, independent of the test assets, of which the measurable quantity is a function. A big question is how to extract the factors from a sample. The difference in the size of the common movements across the test assets is due to sensitivity, also known as asset exposure or loadings, of the test assets to the common factors. Typically, we assume that these exposures are constant over time, but they do not have to be\footnote{Often we would like to determine if the test assets' loadings are time-varying, and then how to model this time-variation. A simple estimation scheme is to sequentially estimate the loadings so we now have a time series of loadings and repeat the process for identifying factor structures in commonly varying time series. If there exists a factor structure in the loadings,  we can model the loadings as $\beta_{i,j,t} \approx a_{i,j} + b_{i,j}Z_t$.}.

Figure \ref{fig: time_series_avg_rv} is suggestive of a common factor structure in firm-level realized variances. This requires us to consider how to best extract factors that explain the common variation in firm-level realized varainces. It also begs the economic question of what causes a firm to have a high level of return volatility on a given day, and is this reason something common to all firms? Recall, return volatility for an average firm appears to go through regimes where large changes are common, and other regimes where small changes are common. Furthermore, when changes in returns are large, they tend to stay large, with similar behavior for for small changes in returns. Such properties are called time-varying volatility and volatility clustering, respectively, and are displayed in \ref{fig: time_series_avg_rv}. There exist many models, typically of the (G)ARCH-type flavor, which try to explain these stylized facts through explicit time series modeling, but a more fundamental question is to explain these stylized facts using economic rationale. \cite{Schwert_why_does_stock_market_vol_change_over_time} looks at possible sources causing time-varying volatility in the S\&P 500 index. He finds little evidence, at the monthly level, that volatility in economic fundamentals such as bond returns, inflation, short-term interest rates, growth in industrial production, and monetary growth, has any observable influence on stock market return volatility. In Schwert's study, much of the movement in stock market return volatility is not explained by the economic variables he examined.

In a first-effort attempt to see if there exists factors that can explain the common time-variation in firm-level volatilities, we use the linear method of principal components to extract common factors.
In many situations, we have a large number of predictor variables that tend to be highly correlated. Commonly, we want to find small numbers of linear combinations of the original predictor variables which are used in place of the predictor variables. Therefore, we seek to find variables that are compressed summaries of the data which captures its \emph{essential} properties. In principal components analysis, a set of variables is approximated with a small number of \emph{underlying factors} that capture a large amount of the common variation among the variables. To construct the principal components used in  our regressions, we use the eigen-decomposition of the firm-level realized variance matrix, 
\begin{equation}
V = Q \Lambda Q^T,
\end{equation}
where $ \Lambda = diag\left(\lambda_1, \lambda_2, \dots,  \lambda_N \right)$ is the diagonal matrix of eigenvalues, ordered in decreasing magnitude, and $Q$ is the matrix of eigenvectors of the firm-level realized variance matrix $V$. \input{sections/PCA} %

Our main model of interest uses this factor extraction methodology and adds the the firm-specific predictive variables from the HAR model in section \ref{corsi_har_subsection},
\begin{equation}
\label{vol_factor model_corsi_regressors_per_firm_per_time}
\hat{\sigma}_{i,t+1}^2 = a_i + \sum_{j=1}^K b_{i,j} \hat{F}_{j,t+1} + \beta_i^{(d)} RV_{i,t}^{(d)} + \beta_i^{(w)}RV_{i,t}^{(w)} + \beta_i^{(m)}RV_{i,t}^{(m)} + \hat{\varepsilon}_{i,t+1}, \quad \forall t, \forall i.
\end{equation}
This variable combination of common factors and firm-specific factors\footnote{This is referred to as model nesting. Specifically the model we consider in \eqref{vol_factor model_corsi_regressors_per_firm_per_time} nests the HAR model and also nests a principal components factor model.} turns out to yield positive forecasting results, as we document in table \ref{model_scoring_statistics}. This is an example of combining common factor and idiosyncratic variables within a single model

There are a few other variants of this methodology that we try, and again, nothing seems to differ too much from this simpler specification \footnote{We consider different numbers of factors, we consider a different factor extraction methodology that accounts for heteroskedasticity in residuals as outlined in \cite{Jones_HFA}, and models that don't include a forecast of residuals, as well.}.

\subsection{Ensemble Models of Realized Variance}

The quality of volatility forecasts depend crucially on a well specified model and the use of \emph{informative} data. An example of a combining models is \cite{Jones_Brandt} where they find a two-factor range-based EGARCH model dominate the extensive set of models and data combinations that they consider, both in-sample and out-of-sample. They combine EGARCH and multi-factor volatility models, while using range-based volatility estimators. They attribute their results to a less misspecified volatility model and a more informative volatility proxy. This indicates volatility forecasts might be improved with other ensembling techniques.

\cite{Diebold_Forecast_Combination} study regularization methods for forecast ensembling. Their methodology, which we replicate in our study, is coined the \emph{egalitarian LASSO} and the \emph{partial-egalitarian LASSO} for forecast combination.
The \emph{egalitarian LASSO} is used for selection and shrinkage towards equal weights. By changing the LASSO penalized estimation problem, we can change the shrinking of weights to zero towards a shrinking of the deviations from equal weights towards zero. eLASSO shrinks in the right direction, however eLASSO selects in the wrong direction. This means that eLASSO tends to not discard forecasters, because it selects and shrinks towards equal weights, rather than zero weights. eLASSO implicitly presumes that all forecasters belong in the set to be combined \footnote{eLASSO can be implemented with a straightforward adaptation of standard LASSO methodology. In particular in  \href{https://www.sas.upenn.edu/~fdiebold/papers2/DieboldShinEgalitarianLasso.pdf}{Appendix A of Diebold and Shin}, we can see to get the eLASSO coefficients, all one needs to do is run the standard LASSO regression on 
$$\tilde{y}_t := y_t - \frac{1}{K} \sum_{i=1}^K \hat{y}_{i, t} \ .$$}. The \emph{partially-egalitarian LASSO} is a modified version of the eLASSO such that some forecasts are potentially discarded, and the survivors are selected and shrunken toward equality. 
The two step implementation, advocated for in \cite[Section 2.6.2]{Diebold_Forecast_Combination}, is first to apply the ordinary LASSO to the set of forecasters to potentially discard some, and second apply eLASSO to the remaining set of forecasters to shrink their weights towards equality.

These procedures can be computationally expensive due to the hyperparamater estimation using cross-validation, and as pointed out in \cite{Diebold_Forecast_Combination}, most of these effectively act as equal weighted averages of all forecasts, or a subset of top-performing forecasts. This insight leads us to consider simple, equal-weighted averages of forecasts from all the previous sections, which has striking success as documented in table \ref{model_scoring_statistics}.

\section{Data Description}
\subsection{Volatility Data}
\label{volatility_data_description}
Herein we describe the construction of our data set for forecasting firm-level realized variances\footnote{All credit for the data set construction goes to Christopher Jones.}. The main variables we construct are daily first and second moments of returns for all firms in the S\&P 500 index from January 1993 to June 2019. To compute the variance of returns, we use the realized variance estimator\footnote{The implementation of the realized variance estimator was motivated by the findings of \cite{patton_beat_5min_rv}.}, which is computed as the sum of all squared 5-minute returns within a single day. To compute the intraday returns on individual firms, we use the NYSE Trades and Quotes (TAQ) database. For every 5-minute interval between $9:30$am and $4$pm ET, we use the last recorded transaction price to be the closing price for that interval. For every firm in the data set, and for every day, there are a total of 84 price observations.

Given these 84 price observations for a single firm on a particular day, we use them to construct 5-minute returns, but before we do that, we have to filter the TAQ observations. In particular, we exclude observations with a price or size of zero, corrected orders, and trades with condition codes B, G, J, K, L, O, T, W, or Z. We include trades on all exchanges, and we also eliminate observations that result in transaction-to-transaction return reversals of 25\% or more, as well as observations that are outside the CRSP daily high/low range. Lastly, we compute size-weighted median prices for all transactions with an identical time stamp. For more on the construction and computing of realized variance using high-frequency data, see \cite{andersen_bollerslev_the_distribution_of_realized_stock_return_volatility}.

\subsection{Options Data}
\label{options_data_description}
For the study of equity option returns, we use the OptionMetrics database\footnote{All credit for the data set construction goes to Christopher Jones.}. OptionMetrics also provides data on interest rates, individual stocks, and equity indices, and more recently borrowing data to analyze short selling. For the study of equity returns, we use the CRSP database, which is the source of stock prices, returns, trading volume, market capitalization, and adjustments for stock splits\footnote{OptionMetrics has a database link-table for the WRDS database, used to merge options and equity data into a single table.} The set of all optionable stocks, i.e. stocks upon which options trade, is much smaller than full set of U.S.-based common stocks that comprise a typical CRSP sample. The OptionMetrics database provides end-of-day bid-ask quotes on all options traded on U.S. exchanges, price, implied volatility\footnote{OptionMetrics interpolates implied volatilities to calculate implied volatilities for options that aren't actually on the market with the correct strike or expiry. They then use the interpolated IVs to calculate option prices and Greeks using the binomial option pricing model, because it allows for early exercise which is common of exchange-traded options.}, and Greeks for all U.S.-listed index and equity options. Our options sample starts in January 1996 and ends in June 2019, which restricts our sample period for the volatility forecasts. We apply various filters to our sample in an effort to remove illiquid securities from the investable universe, while retaining a sufficiently large investable universe to deliver statistically sound results. Many liquidity filters, such as requiring open interest to be nonzero, tend to significantly reduce the sample size, which can cause some issues. 

Our sample consists of call and put options on stocks that are members of the S\&P 500 index. For every stock, we consider the shortest maturity options that have at least ten trading days until expiration and then impose the following filters, where the necessary specification will be subsequently provided:
\begin{itemize}
    \item The bid-ask spread must be less than or equal to half of the midpoint.
    \item The price must be at least 0.1.
    \item Arbitrage bounds must not be violated.
    \item The price cannot exhibit a major reversal.
    \item The quotes must be reasonable.
\end{itemize}
With the above filters applied, we find the call with a delta closest to 0.5, and finding the matching put. Using this call and put, we form a straddle, and compute returns on this straddle using bid-ask midpoints as proxies for the price.

Specifically, we delete observations that violate simple \emph{arbitrage bounds}. We compare the option’s bid or ask to the end-of-day bid or ask on the underlying stock, which is obtained from CRSP. The four different bounds we check are:
\begin{enumerate}
\item For calls, the closing option bid is less than or equal to the closing stock ask price, which prevents the possibility of locking in a guaranteed profit by buying the stock and selling the call.
\item Also for calls, we require that the closing option ask is bounded below the value of immediate exercise, which is computed based on the stock’s closing bid price.
\item For puts, we require that the closing option bid is less than or equal to the strike price, which prevents short put strategies that are guaranteed to have positive instantaneous profits. 
\item Also for puts, we require that the closing option ask is bounded below by the value of immediate exercise, which is computed based on the stock’s closing ask price.
\end{enumerate}

\emph{Reversals} are defined as cases in which option returns, hedged or unhedged, exceed 2000\% (or is below -95\%) and is followed by a similar return below -95\% (or greater than 2000\%).  Such cases are extrememly rare (less than 0.01\%) in the sample, and when they do appear, they seem to be the result of data errors, such as put quotes mistakenly being given for the corresponding call. In these cases, we remove the return on the day of and the day after the apparently incorrect price.

We preserve \emph{reasonable quotes} by deleting what appear to be inaccurate or non-competitive quotes at the start and the end of the holding period.  Specifically, we remove data when the bid price is greater than the ask, or the bid-ask spread is greater than the minimum of \$10 or the stock’s closing price.  While this step only eliminates 0.22\% of the sample, the excluded observations include some highly unrealistic prices, such as an ask price of \$9,999 for a call on a \$40 stock, that have the potential to affect some results.

For more on the use of option filters to construct our sample of options, see \cite{goyal_saretto} and \cite{muravyev_order_flow_expected_option_returns}, and for other subtle remarks regarding construction of option return data sets see \cite[Appendix C, D]{chernov_understanding_index_option_returns}. We should also mention that options can disappear from the investable universe from one period to the next, in-sample, but it is unlikely. 

The construction of the call and put options data set from OptionMetrics allows us to create a subset of data containing S\&P 500 at-the-money, delta-neutral straddles. Such a sub-data set includes for every day $t$, a term structure of implied volatility, for a fixed moneyness, specifically at-the-money. Thus we have the time series evolution of the implied volatility term structure of at-the-money straddles,
\begin{equation}
\left\{IV_t^{\text{ATM-straddle}}(\tau)\right\}_t,
\end{equation}
where for a fixed time $t$, $IV_t^{\text{ATM-straddle}}(\tau)$ is the term structure as a function of time-to-maturity, $\tau$. This subset of data does not contain the implied volatility smile/skew for fixed maturities, because everything is at-the-money, by construction. We need to compute the implied volatility for at-the-money straddles\footnote{Note, this is an instance of the general question of how to compute the implied volatility of a portfolio of options.}. Rather than solving for the Black-Merton-Scholes model's implied volatility using binomial trees, we use an approximation for the implied volatilities 
\begin{equation}
I_{i,t-1}^{\text{strad}} \approx w_{i,t-1}^C I_{i,t-1}^C + w_{i,t-1}^P I_{i,t-1}^P,
\end{equation}
which is the weighted average of the call and put implied volatilities.

Finally, we link the above data description with section \ref{option_portfolio_math}. Our data set contains firm identifiers, calendar-time, strike, time-to-maturity, option (call and put) price, delta, implied volatility, and excess-return computed from the midpoint price, as well as the weights in the call and put, and finally the excess-return of the straddle portfolio.

\section{Results}
\label{forecast_performance_error}
\subsection{Volatility Forecast Rankings}
We conduct a large-scale empirical investigation of realized variance forecasting on a large panel of firms, which to the best of our knowledge, hasn't been done with this magnitude before. We compare the different forecasting models, from section \ref{models_section}, with some standard benchmark models such as rolling standard deviation of returns. We find it is hard to unambiguously and unanimously beat the HAR benchmark model when analyzed through the lens of forecast errors. 

To start, we compute the first four moments of daily stock returns and realized variances. Using our large panel of data, we compute the moments for an average firm, and for an average cross-section of firms. The moments for an average firm were estimated by first computing the moments for a single firm using its time series, and then averaging those firm-specific moments across all stocks in the sample. On the other hand, moments for an average cross-section were estimated by first computing the moments for a cross-section of stocks, and then averaging those cross-section moments over all time periods in the sample\footnote{We should note the firms in the cross-section are changing over time, as our sample is an unbalanced panel.}.

We can think about results corresponding to an average firm and an average cross-section of firms as follows. When one seeks to understand a phenomenon for a general entity, such as individual stocks, bonds, etc., one should use a cross-section average of time series statistics which will summarize the average entity-specific distribution of the relevant variables,
\begin{equation}
\label{average_firm_summary_stat}
\frac{1}{N} \sum_{i=1}^N \hat{\theta}_i (X_{i,1}, \dots, X_{i,T_i}).
\end{equation}
Similarly, when one seeks to understand a cross-sectional phenomenon of entities, one should use a time series average of cross-section statistics that summarize the average cross-section distribution of the relevant variables,
\begin{equation}
\label{average_cross_section_summary_stats}
\frac{1}{T} \sum_{t=1}^T \hat{\theta}_t(X_{1,t}, \dots, X_{N_t, t}).
\end{equation}
In a particular cross-section at time $t$, there are $N_t$-many firms. One may be concerned with the cross-sectional distribution, for reasons such as forming a portfolio of cross-section stocks.

In this study, the relevant variables are firm-level realized variance and returns of all S\&P 500 stocks from 1/1993 to 6/2019. We see in table \ref{summary_stats_variables} the kurtosis of the realized variance of the firms in both the average cross-section and the average firm is large. This tells us we should expect, on an average day, to see firms with large realized variances relative to other firms in the cross-section. That is, there are firms that dominate the cross-section in terms of their realized variance. The kurtosis for the average firm is also large, which tells us we should expect an average firm in the S\&P 500 index to have large realized variance outliers over the course of its time series.

\input{tables/regression_variables_summary_stats}

Table \ref{summary_stats_variables} says there are days where particular firms display large spikes in realized variances relative a firm's own history, as well as relative to other firms in a cross-section. 

We move to forecasting realized variance using the variables summarized in table \ref{summary_stats_variables}. To start, we first look at a measure of historical realized variance, which is a moving average of squared full-day returns, which we call rolling standard deviation squared. Next, we test Corsi's HAR model of realized variance, as described in \ref{corsi_har_subsection}. Such a model has a self-imposed sparsity structure on the predictor variables and is simple to implement on a rolling basis with three predictors and 250 observations. We take the rolling standard deviation squared and the HAR model to be our benchmarks. Subsequently, we investigate the ability of the class of \emph{naive} penalized high-dimensional regression models to forecast daily firm-level realized variance. We consider these to be naive because we add all lagged realized variances and lagged returns over daily, weekly, and monthly horizons of all firms in the cross-section for the entire estimation sample. Every penalized regression forecast has approximately 6,000 regressors with 250 observations in the estimation sample. 

The common time-varying nature of average firm-level realized variances, as shown in \ref{fig: time_series_avg_rv}, is suggestive of a common factor structure in daily firm-level realized variances. Accordingly, we use principal components methodology to extract sources of common variation of realized variance across firms. We then use the top three principal components as well as the predictor variables for the HAR baseline model to form a \emph{nested} model which forecasts next-day realized variance. Finally, motivated by \cite{Diebold_Forecast_Combination}, we consider simple equal-weighted averages of our forecasting models. This simple version of forecast ensembling is known to perform well out-of-sample, though not necessarily optimally, as documented in \cite[Section 4]{Diebold_Forecast_Combination}, yet have the benefit of not having to train additional models or search for optimal hyperparameter configurations.

\input{tables/model_scoring_stats}

We see in table \ref{model_scoring_statistics} the results are not unanimous across panels and across error measurements. In panel A, the benchmark HAR has the smallest loss function error, for RMSE, MAE, and QLIKE\footnote{QLIKE is less sensitive to large observations, relative to RMSE, by more heavily penalizing under-predictions than over-predictions.} loss functions, with the LASSO high-dimensional penalized regression coming in a close second. However, the $R^2$ of the MZ-regression, as described in \ref{vol_oos_performance_measurement}, is largest for the simple ensemble model, with HAR coming in a close second. In panel B, we see a similar story, namely, the benchmark HAR model minimizes the loss function error with the LASSO model coming in an even tighter second place. Again, as in panel A, the $R^2$ of the MZ-regression ranks HAR and forecast averaging in the top two, but this time the ranking is reversed with HAR beating forecast averaging. Finally, in panel C, we see the most diverse set of rankings. According to RMSE, the rolling standard deviation squared beats HAR and the LASSO, where this time the LASSO loss is lower than the HAR loss. We see, however, according to MAE and QLIKE, panel C agrees with panel A and B with HAR minimizing loss and LASSO coming in a close second. The nested PCA-HAR model as well as forecast averaging perform the best according the the $R^2$ of the MZ-regression. As previously noted, the presence of large outliers and noise in the proxy for realized variance likely contributed to the lack of consensus of forecast rankings across all panels and error measurements. 

Though rarely beating the HAR benchmark, it is important to notice the LASSO forecast does a good job for being so naive. This is impressive due to the large number of highly correlated predictors relative to the small number of observations, as well as the naive ``throw-everything-in'' approach to modeling firm-level volatility. This points to the possibility of greatly improving a volatility forecasting model using more tailored penalized regression techniques. Issues needing to be more deeply considered when using out-of-the-box high-dimensional penalized regressions are how to handle collinearity and scaling of the predictor variables. Moreover, it would be a worthwhile pursuit to consider customized penalty functions which encode economically motivated priors that can help steer the model towards capturing the stylized facts in volatility\footnote{Including clustering, leptokurtic distributions, asymmetry and the leverage effect, as well as response to external shock events as documented in \cite{patton_what_good_is_a_volatility_model}.}. We see the deviations in forecast ranking when using loss functions, which tend to favor HAR and LASSO, as well as using the $R^2$ of MZ-regressions, which tend to favor the  nested PCA-HAR model and the equal-weighted forecast averaging model.

Though the results are the least unanimous in panel C, we believe panel C's methodology is the most well-suited for aggregating the large panel of forecast errors\footnote{Pooled forecast error corresponds to 
\begin{equation}
\frac{1}{NT}\sum_{i=1}^{N}\sum_{t=1}^{T_i} \widehat{\varepsilon}_{i,t}
\end{equation}
where $N$ and $T$ are the total number of firms and days in the entire sample, respectively, and $T_i$ is the number of days in the $i^{th}$ firm's time series.} because to perform our out-of-sample walk-forward validation, as described in section \ref{oos_analysis}, we re-estimate our forecasts every day, for every firm. In the full panel of data, we model each entry (day and time) individually, and panel C's measurement of errors averages each entry of errors (day and time) equally. Specifically, we do not model a single firm for all time, nor do we model an entire cross-section at each point in time, which corresponds to panel A and B's measurement of errors.

There remains open questions as to how to properly interpret these results. This is no surprise. The ambiguity of out-of-sample volatility forecast error measurement was documented in \cite{patton_vol_loss_functions_2011}, who favored mean-squared error and QLIKE, as opposed to \cite{Meddahi_justification_for_MZ_regression} who favored the $R^2$ of MZ-regressions. To further add to the ambiguity,  marginal improvements to out-of-sample statistical measures of forecast error can lead to economically large improvements in portfolio performance, as observed for the first moment of returns in \cite{campbell_predicting_oos_can_anything_beat_naive_benchmarks}, and is observed in our sample as well, as documented in \ref{option_return_forecast_results}. We believe these models collectively provide favorable evidence of utilizing techniques such as model ensembling, factor models of firm volatility, as well as high-dimensional penalized regressions to forecast volatility, which breaks the $\mathbb{P}-$measure tradition of classical (G)ARCH-type volatility forecasting which is well-studied in the literature. We have chosen rather simple and off-the-shelf implementations of ensembling, factor, and penalized regression models. This begs the question of investigating these classes of models with implementations more specifically tailored towards stylized facts of volatility. Further work should be done to enhance the different model classes to exploit their full potential and see how much improvement they can have over traditional methods of volatility modeling and forecasting. One particular question that also warrants further attention is how to make sense of the outliers? That is, as documented in table \ref{summary_stats_variables}, the kurtosis is extremely large and these outliers are important for predicting variance\footnote{If we miss the outliers, then we have large forecast errors; also note that large $R^2$s are frequently caused by a single outlier.}. It remains to be seen if outliers aren't necessarily as important for predicting option returns, as in section \ref{option_return_forecast_results}, as they are for predicting volatility. This leads us to a conjecture: an outlier in volatility forecasts isn't an outlier in the spread between realized voaltility and implied volatility, which is the variable used for predicting option returns. That is, when realized volatility is large, it is likely implied volatility is also large, so the sorting variable doesn't have as much kurtosis as volatility itself, which we observe in panel C of table \ref{options_summary_stats_variables}.

\subsection{Option Portfolio Returns}
\label{option_return_forecast_results}

In many machine learning and forecasting applications, the objective for estimating a model's parameters, in-sample, is to minimize the sum of squared forecast errors or functions of it, such as $R^2$. The corresponding sum of squared, out-of-sample, prediction errors is used to measure the out-of-sample predictive performance. This method of minimizing prediction error is also employed in the context of variance or volatility forecasting. As noted in \cite[Chapter 3.2]{nagel_ml_in_asset_pricing} it is not obvious that this is the right approach to evaluate the power of a forecast. In a trading or asset management setting, the ultimate goal is not to forecast moments of returns, but rather to use forecasts to construct portfolios that earn large out-of-sample expected returns, either absolute or risk-adjusted. Methods that lead to better forecasts in terms of standard loss functions do not necessarily lead to better portfolios in terms of standard portfolio metrics. Similarly, methods that lead to better portfolios may not necessarily be related to better forecasts, and certainly not in the same magnitudes.

We first consider summary statistics of the unconditional average excess returns of an equally-weighted at-the-money equity straddle portfolio. As documented in panel A of table \ref{options_summary_stats_variables}, the average unconditional excess straddle returns are slightly negative, at approximately negative half of a basis point, have positive skewness, and have a high degree of excess kurtosis. In panel B of table \ref{options_summary_stats_variables}, we see the average implied volatility of at-the-money straddle returns is much larger than realized volatility, and far less skewed. We also notice the average realized volatility coming from the forecasts are significantly lower than the measured realized volatility, with a significant increase in skewness and kurtosis, with the LASSO and HAR models showing the most dramatic increase skewness and kurtosis. Panel C in table \ref{options_summary_stats_variables} computes summary statistics for the sorting variable, typically named the volatility risk premium,
\begin{equation}
\ln\left(\frac{\widehat{RV}}{IV}\right)
\end{equation}
for each forecast. The sorting variable's average value is positive for the PCA-HAR nested model and the forecast averaging model, and negative for HAR, LASSO, and rolling standard deviation, with a skewness much closer to zero, and kurtosis\footnote{We're using excess kurtosis, meaning a standard normal distribution has an excess kurtosis of zero.} much lower than the realized variance forecasts.

\input{tables/options_sample_summary_stats}

In table \ref{portfolio_stats_sharpe}, we find economically significant unannualized average returns ranging from $1\%$ to $2\%$ per day, and unannualized daily Sharpe ratios ranging from $0.57$ to $0.80$. This confirms the so-called volatility risk premium is large and economically significant in daily equity options returns, which adds to the existing literature of \cite{goyal_saretto} who document such returns at the monthly horizon.

Given the problems outlined in section \ref{vol_oos_performance_measurement} with using traditional forecast error measurements, we use the conditional straddle portfolio performance measures, in table \ref{portfolio_stats_sharpe}, as a different way to rank volatility forecasts without reference to a particular loss function. We are particularly interested in the high minus low portfolio, that is, the portfolio that longs the highest quintile of stocks and shorts the stocks in the lowest quintile\footnote{We can think of these long-short portfolios as ``market neutral'' because the portfolio weights will sum to zero.}. All portfolios are formed by equally weighting stocks in the portfolio.

\input{tables/portfolio_stats_sharpe_mean}

We observe the forecasts distinguish themselves in economically significant average returns on long-short quintile portfolios, on an absolute and risk adjusted basis. These portfolios are formed on the spread between forecasted realized and implied volatility. The PCA-HAR nested model, from section \ref{Low-Dimensional Statistical Factor Models of Realized Variance}, separates stocks in the cross-section to form long-short straddle portfolios that outperform all other volatility forecasting models. It beats the benchmark by increasing the Sharpe ratio by $20\%$ as well as improving by average daily return by approximately $0.6\%$ (or $60$ basis points). The equally-weighted average forecast ensemble has similar improvements over the benchmark. While the portfolios formed from the LASSO model have lower daily average returns, the model also produces a drop in portfolio return volatility. This is evidenced by the LASSO-based long-short portfolio's Sharpe ratio, which performs similar to the rolling standard-deviation benchmark model. The results in table \ref{portfolio_stats_sharpe} are suggestive of the potential value of combining common factor and idiosyncratic variables within a single model.

These results strengthen the case of equity option return predictability at the daily frequency, based on the spread between realized and option implied volatility. There is more work to be done here in making these results more pointed both for understanding forecast evaluation, and for crafting a true trading strategy, in particular understanding the effects of transaction costs and margin requirements as documented in \cite{goyal_saretto}. 

One thing to note is the moments in the conditional and unconditional straddle portfolios, as in tables \ref{summary_stats_straddle_returns} and \ref{options_summary_stats_variables}, respectively. Particularly, when compared to the unconditional straddle portfolios, the conditional straddle portfolios display an increase in mean, a decrease in standard deviation, a decrease in skewness, and an especially large decrease kurtosis. This suggests the conditional returns are brought closer a positively biased normal distribution. In table \ref{summary_stats_straddle_returns}, we particularly note the increase in means from the baseline models to top-performing models is associated with an increase in skewness and kurtosis. 

\input{tables/top_performing_portfolio_summary_stats}

It should be noted that the returns highlighted in table \ref{portfolio_stats_sharpe} are not entirely representative of the actual returns that could be realized by a long-short straddle portfolio. This is due to transaction costs and margin requirements when short selling straddles, as mentioned in \cite[Section 5.2]{goyal_saretto}. Further, option returns have a bias, especially at higher frequencies, as shown in  \cite{jones_noisy_option_returns}. We conjecture such microstructure biases won't affect our results too much because we're working with at-the-money options, and microstructure biases tend to be larger in options that are away-from-the-money.

A deeper analysis on transaction costs, or how to optimally implement the above edge through transaction cost and order type analysis is certainly warranted. In the literature, transaction costs are typically estimated by estimating the bid-ask spread and using this measure as a proxy for transaction costs, which will be an upward biased estimate of the true cost one would incur while trading. It is important to recognize bid-ask spreads are not constant, and in practice one's goal is to trade when the spread is small, which would require additional modeling of the dynamics of the spread. Also, the use of order types other than a market order, such as a limit order or more complex order types, can  help reduce trading costs. And finally, the signal we investigated above, the discrepancy between realized and implied volatility, is not intended to be used as a stand-alone strategy, but as an additional signal meant to be used in a complex trading program combining multiple signals. Though we believe transaction costs must be analyzed, the above reasons mildly critique the literature's method of conducting transaction cost analysis, and suggest a list of questions for further research. 

With regards to models, the results in section \ref{forecast_performance_error} point to the class of high-dimensional penalized regression as having the greatest potential for minimizing forecast errors, and from results in this section \ref{option_return_forecast_results} point to the class of volatility factor models as having the greatest potential to improve option portfolio performance. Further research needs to investigate this discrepancy between the volatility forecast error measurements and the standard portfolio performance metrics. Such discrepancies raise questions regarding the best way to train machine learning models for forecasting volatility. In particular, should we be using classical error measurements to evaluate the power of an out-of-sample volatility forecast? Also, should we be using the standard maximization of the cross-validated $R^2$ to tune model hyperparameters? This is left for future research.

\section{Conclusion}

In this paper we focused on finding a variance forecasting methodology which beats the de-facto benchmark of \cite{corsi_har_rv_model}. We examined high-dimensional penalized regressions, such as LASSO, low-dimensional statistical factor models, as well as simple ensembling methods, and found we can beat the benchmark, though marginally and not unanimously according to different measures of forecast errors. We found, however, the forecasts, $\widehat{\sigma}^{\mathbb{P}}$, do distinguish themselves in the context of equity option returns, and provide economically significant improvements to the benchmark. One distinction of our investigation is we forecasted individual firm-level return variance at the daily frequency for all firms in the S\&P 500, over our twenty-five year sample. To our knowledge, no study of this magnitude has been done in the literature before.

We then looked at volatility from the perspective of an end-user of volatility forecasts. Inspired by the work of \cite{goyal_saretto}, we considered a high-frequency, daily, at-the-money straddle return strategy which captures discrepancies between our forecasts of volatility and the \emph{market's} forecast of volatility, $\widehat{\sigma}^{\mathbb{Q}}$. Strictly speaking, we considered the log difference of realized volatility and Black-Scholes at-the-money implied volatility as a predictor variable for straddles. These portfolios are capturing the signal between a measure of the spread between the second moments of the $\mathbb{P}$ and $\mathbb{Q}$ distribution of returns, 
\begin{equation}
\ln\left(\frac{\sigma^{\mathbb{P}}}{\sigma^{\mathbb{Q}}}\right).
\end{equation}
We observed economically significant returns, both absolute and risk-adjusted, that distinguish our above forecasts, favoring the nested factor model and simple ensembling models. 

Our results give rise to a few important directions in volatility modeling as well as understanding the role of forecasts from option market participants. First, we observe a very naive high-dimensional model can match the benchmark models with no additional modeling effort. Thus incorporating economic restrictions into the modeling framework for high-dimensional machine learning techniques will likely lead to a fruitful increase in volatility forecasting. Secondly, due to the large comovement of firm-level variances and the out-performance of the nested factor model, it will be useful to understand what information common factors of firm-level variances contain that isn't being accounted for in the baseline model. Finally, because of the ambiguity present in traditional forecast rankings, the question remains of why we choose to estimate models with traditional loss functions, rather that a loss function designed around portfolio performance.

\clearpage

\clearpage

\appendix
\section{Appendix: Simple Option Portfolio Math}
\label{option_portfolio_math}

In this section, we digress for the interests of the mathematically inclined reader to add some formalism. We consider options, calls or puts, and denote the price as
\begin{equation} 
\label{option_price_full}
\mathcal{O}\left(t, S_t, I_t; K, T\right),
\end{equation}
where at time $t$, the underlying spot price is $S_t$, with an implied volatility of $I_t$, a strike of $K$ and an expiration at time $T$ \footnote{We can think of the $(K,T)$-plane as the space of all options written, at time $t$, on the underlying $S_t$. In other words, the $(K,T)$-plane is the investable universe of options on an underlying at a fixed point in time.}. 

We will be selecting from this investable universe of options, i.e. choosing options from the $(K,T)$-plane, based on properties such as time-to-maturity, $\tau = T-t$, moneyness, $\frac{S_t}{K}$, value of implied volatility, $I_t$ \footnote{Specifically the spread between implied volatility and the underlying's realized volatility.}, as well as possibly other risk measures such as the delta and other Greeks. Upon the appropriate selection of options from the investable $(K,T)$ universe, we will weight the options and add them together to form an option portfolio.

A portfolio is a linear combination of securities. The change in dollar value of a portfolio at time $t$ is defined as 
\begin{equation}
\label{portfolio_change_in_value_general}
V_t^p - V_{t-1}^p := \sum_{i=1}^N n_{i,t-1}(S_{i,t} - S_{i,t-1}),
\end{equation}
where $n_{i,t-1}$ is the number of shares of the $i^{th}$ security, set at time $t-1$, and $S_{i,t}$ is the price of the $i^{th}$ security observed at time $t$, for all $N$-many securities in the portfolio. We can also call \eqref{portfolio_change_in_value_general} the PnL (profit-and-loss) of the portfolio, and it can be interpreted as the return per dollar value invested\footnote{A \emph{return} is, most generally, a payoff divided by a unit of some price, where this price doesn't have to be the initial price or value of the portfolio. That is, a \emph{return} can be thought of as the dollar payoff (PnL) of a portfolio per unit of some other variable, typically a price.}. The change in portfolio value from time $t-1$ to $t$ comes from the change in security prices since the number of shares remains fixed. 

We can convert this to a portfolio rate of return  
\begin{equation}
\label{portfolio_return_general}
R_t^p = \frac{V_t^p - V_{t-1}^p}{V_{t-1}^p} = \sum_{i=1}^N \left(\frac{n_{i,t-1} S_{i,t-1}}{V_{t-1}^p}\right)\left(\frac{S_{i,t} - S_{i,t-1}}{S_{i,t-1}}\right) = \sum_{i=1}^N w_{i,t-1} R_{i,t},
\end{equation}
where we define the portfolio weights and security rate of returns as
\begin{align*}
    w_{i,t-1} & := \frac{n_{i,t-1} S_{i,t-1}}{V_{t-1}^p}, \\
    R_{i,t} & := \frac{S_{i,t} - S_{i,t-1}}{S_{i,t-1}}.
\end{align*}
Finally, we can define the portfolio returns in excess of the risk free rate 
\begin{equation}
\label{excess_portfolio_return_general}
R_t^{e,p} := R_t^p - R_t^{rf} = \sum_{i=1}^N w_{i,t-1}\left(R_{i,t} - R_{t}^f\right).
\end{equation}

We now consider how \eqref{portfolio_change_in_value_general}, \eqref{portfolio_return_general}, and \eqref{excess_portfolio_return_general} change in the context of specific portfolios of options. For simplicity, consider a vanilla option, either a call or put, written on an underlying security, say a stock. We denote the price of the option and its underlying at a particular time by $\mathcal{O}_{i,t}^{(K,T)}$, as short-hand for \eqref{option_price_full}, and $S_{i,t}$, respectively. Though simple, we highlight the fact that a delta-hedged option is, in fact, a portfolio. This portfolio consists of a long position in an option and a short position in its underlying in the amount delta-many shares. The change in dollar value \eqref{portfolio_change_in_value_general} of a delta-hedged portfolio at time $t$ is 
\begin{equation}
\label{delta_hedged_portfolio_value_change}
    V_t^{\substack{\text{delta} \\\text{hedged}}} - V_{t-1}^{\substack{\text{delta} \\ \text{hedged}}} = (\mathcal{O}_{i,t}^{(K,T-1)} - \mathcal{O}_{i,t-1}^{(K,T)}) - \Delta_{i,t-1}^{\mathcal{O}}(S_{i,t} - S_{i,t-1}),
\end{equation}
where the option's delta is defined as
\begin{equation}
    \Delta_{i,t-1}^{\mathcal{O}} := \frac{\partial \mathcal{O}_{i,t-1}^{(K,T)}}{\partial S_{i,t-1}} (t-1, S_{i,t-1})
\end{equation}
and measures the option's price sensitivity to small changes in the underlying spot price. It is important to realize that the although the delta is a function of spot price and time, it is measured at time $t-1$, and thus a constant at time $t$, not a function. At $t-1$ this portfolio is formed to have zero delta, but at time $t$, you are effectively holding a different option because the time to maturity has decreased. That is, a $T-$day option at time $t-1$ is now a $(T-1)-$day option at time $t$ with a different delta. This allows us to compute the option portfolio's new delta at time $t$, 
\begin{equation}
   \frac{\partial V_t^{\substack{\text{delta} \\\text{hedged}}}}{\partial S_{i,t}} = \frac{\partial\left(\mathcal{O}_{i,t}^{(K,T-1)} - \Delta_{i,t-1}^{\mathcal{O}}S_{i,t}\right)}{\partial S_{i,t}} = \Delta_{i,t}^{\mathcal{O}} - \Delta_{i,t-1}^{\mathcal{O}}. 
\end{equation}

Now that we know the change in value of a delta-hedged portfolio \eqref{delta_hedged_portfolio_value_change}, we can compute the excess rate of return \eqref{excess_portfolio_return_general} of a delta-hedged option portfolio by 
\begin{equation}
    R_t^{e, \ \substack{\text{delta} \\ \text{hedged}}} = \frac{\mathcal{O}_{i,t-1}^{(K,T)}}{V_{t-1}^{\substack{\text{delta} \\ \text{hedged}}}} \left(\frac{\mathcal{O}_{i,t}^{(K,T-1)} - \mathcal{O}_{i,t-1}^{(K,T)}}{\mathcal{O}_{i,t-1}^{(K,T)}} - R_{t-1}^{rf}\right) - \frac{\Delta_{i,t-1}^{\mathcal{O}} S_{i,t-1}}{V_{t-1}^{\substack{\text{delta} \\ \text{hedged}}}} \left(\frac{S_{i,t}-S_{i,t-1}}{S_{i,t-1}} - R_{t-1}^{rf}\right), 
\end{equation}
where $R_{t-1}^{rf}$ is the product of the riskless rate of return available at time $t-1$ and the number of calendar days between time $t-1$ and $t$, since we know interest accrues on a calendar time basis. 

Given we now understand how to connect familiar notions of option prices and deltas to portfolios and returns, we can now begin to think about options from an investment point-of-view ($\mathbb{P}$-measure), rather than a pricing and hedging perspective ($\mathbb{Q}$-measure). In this study, we're interested in straddle returns, and in section \ref{options_data_description} we describe how we compute the straddle returns from our data set. More specifically, we're interested in at-the-money, delta-neutral straddles. For the record, a straddle is a different type of portfolio of options where you go long, buy, a call and put option on the same underlying security with the same strike and expiration date. Such a portfolio has the property, by definition, that exactly one \emph{leg}, i.e. one option in the portfolio, will expire in-the-money as long as the terminal price differs from the spot price at the time of the portfolio formation. Again, this is due to the fact that we're only considering at-the-money straddles, so the strike of the call and put equals the spot price, and if the spot differs from the price at expiry, then the price at expiry differs from the strike, and exactly one of the straddle's legs will be in-the-money.

To construct such straddle portfolios, rather than trading in the underlying to create a delta-neutral portfolio, for each option in the portfolio, we don't need to trade in the underlying and we can just rebalance our options portfolio. Rather than buying a call and a put and then delta-hedging in the underlying, you can just rebalance the portfolio weights. That is, we simply use the weights in the call and put to make the straddle delta-neutral.

Analogous to \eqref{option_price_full}, let $C_{i,t}^{(K,T)}, \ P_{i,t}^{(K,T)}$ be the price of a call and put option, respectively, on the $i^{th}$ underlying with spot value of $S_{i,t}$, at time $t$, with an implied volatility $I_t^C(K,T)$ and $I_t^P(K,T)$, for a fixed strike $K$ and expiry $T$. Then the value of the straddle portfolio can be written as
\begin{equation}
\label{straddle_portfolio_pnl}
V^{\text{strad}}_{i,t} = n_{i,t-1}^C C_{i,t}^{(K,T)} + n_{i,t-1}^P P_{i,t}^{(K,T)}.
\end{equation}

Equivalently, we see the change in portfolio value is due to changes in the assets in the portfolio
\begin{equation} 
\label{straddle_portfolio_chainge_in_pnl}
V^{\text{strad}}_{i,t} - V^{\text{strad}}_{i,t-1} = n_{i,t-1}^C \left( C_{i,t}^{(K,T-1)} - C_{i,t-1}^{(K,T)} \right) + n_{i,t-1}^P \left(P_{i,t}^{(K,T-1)} - P_{i,t-1}^{(K,T)}\right).
\end{equation}
Dividing by the total value of the portfolio and cleverly multiplying by one, we can express the dollar value of the straddle portfolio in terms of returns
\begin{equation}
\label{straddle_portfolio_return}
R_t^{\text{strad}} = \frac{V^{\text{strad}}_{i,t} - V^{\text{strad}}_{i,t-1}}{V^{\text{strad}}_{i,t-1}} = \left(\frac{n_{i,t-1}^C C_{i,t-1}^{(T,K)}}{V^{\text{strad}}_{i,t-1}}\right) \frac{\left( C_{i,t}^{(K,T-1)} - C_{i,t-1}^{(K,T)} \right)}{C_{i,t-1}^{(K,T)}} + \left(\frac{n_{i,t-1}^P P_{i,t-1}}{V^{\text{strad}}_{i,t-1}}\right) \frac{\left(P_{i,t}^{(K,T-1)} - P_{i,t-1}^{(K,T)}\right)}{P_{i,t-1}^{(T,K)}},
\end{equation}
which can be written as
\begin{equation} 
\label{straddle_portfolio_simplified_return}
R_{i,t}^{\text{strad}} = w_{i,t-1}^C R_{i,t}^C + w_{i,t-1}^P R_{i,t}^P.
\end{equation}
The delta of the straddle is the sensitivity of the portfolio with respect to a change in spot
\begin{equation} 
\label{straddle_portfolio_delta}
\Delta^{\text{strad}}_{i,t} = \frac{\partial V^{\text{strad}}_{i,t}}{\partial S_{i,t}} \approx \frac{V^{\text{strad}}_{i,t} - V^{\text{strad}}_{i,t-1}}{S_{i,t} - S_{i,t-1}},
\end{equation}
where the numerator is calculated in \eqref{straddle_portfolio_chainge_in_pnl}. Using \eqref{straddle_portfolio_pnl}, we can calculate the straddle delta in terms of the delta of the portfolio's call and put, as well as the number of shares in each asset 
\begin{equation}
\label{}
\Delta^{\text{strad}}_{i,t} = n_{i,t-1}^C \Delta^{C}_{i,t} + n_{i,t-1}^P \Delta^{P}_{i,t},
\end{equation}
which shows we can control the delta of the straddle by choosing a portfolio of calls and puts weighted in the right amount. We will now show what the \emph{right amount} is, and because we only care about the weights of the straddle portfolio, we can assume, without loss of generality, we'll always hold one call option which allows us to only have to find the number of puts required to make the straddle portfolio delta-neutral. Assume
\begin{equation}
n_{i,t-1}^C = 1, \text{ then } \Delta^{\text{strad}}_{i,t-1} = \Delta^{C}_{i,t-1} + n_{i,t-1}^P \Delta^{P}_{i,t-1} = 0
\end{equation}
which implies
\begin{equation}
\left(n_{i,t-1}^C, n_{i,t-1}^P \right) = \left(1, -\frac{\Delta^{C}_{i,t-1}}{\Delta^{P}_{i,t-1}}\right).
\end{equation}
Given the number of puts and number of calls, we can then determine the weights for a delta-neutral straddle by 
\begin{equation}
\label{weight_call_straddle_delta_neutral}
w_{i,t-1}^C = \left(\frac{n_{i,t-1}^C C_{i,t-1}}{V^{\text{strad}}_{i,t-1}}\right) = \left(\frac{1 C_{i,t-1}}{V^{\text{strad}}_{i,t-1}}\right),
\end{equation}
and
\begin{equation}
\label{weight_put_straddle_delta_neutral}
w_{i,t-1}^P = \left(\frac{n_{i,t-1}^P P_{i,t-1}}{V^{\text{strad}}_{i,t-1}}\right) = \left(\frac{-\frac{\Delta^{C}_{i,t-1}}{\Delta^{P}_{i,t-1}} P_{i,t-1}}{V^{\text{strad}}_{i,t-1}}\right).
\end{equation}

Computing the weights is further complicated because there isn't a unique \emph{price} for the calls and puts, but rather a bid-ask spread. Given we know quotes at which people are willing to buy and sell, for particular volumes,  we must decide what quote, or function of quotes, we should choose as a proxy for price when computing returns. A common, though not necessarily realistic, estimate of a fill price is to take the midpoint price, that is the transaction price where no bid-ask spread is paid. In other words, one can only observe a price when a transaction occurs; if no transaction has occurred, then we can use the midpoint of the bid and ask quotes as a proxy for a price,
\begin{equation}
\label{midpoint_price}
\mathcal{O}_{i,t}^{\text{mid}} = \frac{1}{2}\mathcal{O}_{i,t}^{\text{bid}} + \frac{1}{2}\mathcal{O}_{i,t}^{\text{ask}}
\end{equation}
which is an average of the highest bid and the lowest ask price\footnote{Note this could be refined by taking a weighted average of all quotes in the limit order book, where the weights are a function of the volume.}. 

Finally, given we can now compute the weights for a delta-neutral straddle portfolio for every straddle in the investable universe $(K,T)$, we must narrow down all of the straddles to those straddle portfolios which are at-the-money. Thus we have a mathematical sketch of how to create a data set of delta-neutral, at-the-money straddle portfolio returns, further details are given in \ref{options_data_description}.

\clearpage
\bibliographystyle{abbrvnat}
\bibliography{bibliography/ref}

\end{document}

%% file: sections/PCA.tex
More precisely, again, due to the large degree of common movement of firms' total variance over time, which is suggestive of a common factor structure in firm specific total variance, we believe 
\begin{equation}
\label{vol_factor model_per_firm_per_time}
\sigma_{i,t}^2 = a_i + \sum_{j=1}^K b_{i,j}F_{j,t} + \varepsilon_{i,t}, \quad \forall t, \forall i,
\end{equation}
for some set of common factors $\{F_j\}_{1\leq j \leq K}$. Believing firm-level realized variance is a linear function of contemporaneous factors at each point in time, allows us to express \eqref{vol_factor model_per_firm_per_time} in cross-sectional form as
\begin{equation}
\label{vol_factor_model_per_time}
\mathbf{V}_{N\times 1}(t) = \mathbf{a}_{N\times 1} + \mathbf{B}_{N\times K}\mathbf{F}_{K\times 1}(t) + \mathbf{\varepsilon}_{N\times 1}(t), \quad \forall t,
\end{equation}
where the factors are the variables that best explain the historical variability of the cross-section of realized volatilities, $\mathbf{V}$. We can then use ordinary least squares to estimate $\mathbf{a}$, $\mathbf{B}$, and $\mathbf{\varepsilon}$. We could consider stacking the above vector equation in time, so we have a $T \times N$ matrix of realized variances; it is useful to stack and think in terms of matrices and linear algebra when considering projections and rotations of data, but this isn't always the best way to think of these models in a rolling time-series regression.

Our goal is to predict $\mathbb{E}_{t}\left[\mathbf{V(t+1)}\right]$. If we assume firm-level realized variances are linear combinations of common factors as in \eqref{vol_factor_model_per_time} and take conditional expectations, we get
\begin{equation}
\label{variance_forecast}
\widehat{\mathbf{V}}(t+1) := \underbrace{\mathbb{E}_{t}\left[\mathbf{V}(t+1)\right]}_{\substack{\text{forecast for} \\ \text{firm-level} \\ \text{realized variances}}} = \mathbf{a} + \mathbf{B}\underbrace{\mathbb{E}_{t}\left[\mathbf{F}(t+1)\right]}_{\substack{\text{forecast for} \\ \text{factor values}}} + \underbrace{\mathbb{E}_{t}\left[\mathbf{\varepsilon}(t+1)\right]}_{\substack{\text{forecast for} \\ \text{residual, which isn't} \\ \text{necessarily zero}}},
\end{equation}
which necessitates the need to forecast the conditional expectation of future factor levels and residuals if we hope to forecast the firm-level realized variances. We forecast the conditional expected future factor levels as the linear combination of average lagged factor levels over daily, weekly, and monthly horizons
\begin{equation}
\label{factor_forecast}
\mathbf{\hat{F}}(t+1) := \mathbb{E}_{t}\left[\mathbf{F}(t+1)\right] = \mathbf{\tilde{a}} + \mathbf{\tilde{b}}_1 \mathbf{F}(t) + \mathbf{\tilde{b}}_2 \mathbf{\bar{F}}(t-5, t) + \mathbf{\tilde{b}}_3 \mathbf{\bar{F}}(t-22,t) \footnote{\text{If we believed factors are particularly persistent, we could forecast the factors as the lagged values $\mathbf{\hat{F}}(t+1) = \mathbf{F}(t)$.}}
\end{equation}
Similarly, we can model the noise as
$$
\mathbf{\varepsilon}(t+1) = \mathbf{\phi}_0 + \mathbf{\phi}_1 \mathbf{\varepsilon}(t) + \mathbf{\phi}_2 \mathbf{\bar{\varepsilon}}(t-5, t) + \mathbf{\phi}_3 \mathbf{\bar{\varepsilon}}(t-22,t) + \mathbf{\eta}(t+1),
$$
which implies a forecast of the conditional expected future residual of
\begin{equation}
\label{residual_forecast}
\mathbf{ \hat{\varepsilon}}(t+1) = \mathbb{E}_{t}\left[\mathbf{\varepsilon}(t+1)\right] = \mathbf{\phi}_0 + \mathbf{\phi}_1 \mathbf{\varepsilon}(t) + \mathbf{\phi}_2 \mathbf{\bar{\varepsilon}}(t-5, t) + \mathbf{\phi}_3 \mathbf{\bar{\varepsilon}}(t-22,t).
\end{equation}
Plugging in \eqref{factor_forecast} and \eqref{residual_forecast} into \eqref{variance_forecast} gives us a forecast for firm-level realized variances.

We do this factor extraction and forecasting process on a rolling daily basis. Every day, we extract the factors from the cross-section of firm-level realized variances over the previous year. These factors are driving the variability of firm-level realized variance over that year; we use these factors to predict the next-day realized variance of all firms in the cross section. 
 

%% file: tables/regression_variables_summary_stats.tex
\begin{table}[htb]
\centering
\ra{1.3}
\caption[Regression variables summary statistics.]{Daily summary statistics of the first and second moments of stock returns. That is, we look at the first four moments of stock returns and realized variances. Panel A gives summary statistics for an average firm in our sample. Panel B gives summary statistics for an average cross-section of firms. We see realized variance is prone to large outliers both cross-sectionally and firm-specifically.}
\label{summary_stats_variables}
\begin{tabular}{lllll}
\hline
Variable & Mean & SD & Skew & Kurtosis\\
\hline
\hline
\multicolumn{5}{c}{\textbf{Panel A: Average Firm}}\\%
\hline
Return &  0.00028 &  0.01605 &  0.136 & 1.64 \\
Realized Variance & 0.00045 &  0.00072 & 0.312 & 71.2 \\
\hline
\hline
\multicolumn{5}{c}{\textbf{Panel B: Average Cross-Section of Firms}}\\%
\hline
Return &  0.00009 &  0.02054 &  0.155 &  4.41 \\
Realized Variance &  0.00057 &  0.00096 & 9.68 & 258 \\
\hline
\end{tabular}
\end{table}

%% file: tables/model_scoring_stats.tex
\begin{table}[htb]
\centering
\ra{1.3}
\caption[Forecast error measurements.]{Model scoring statistics using four common scoring functions: root-mean-squared-error, mean-absolute-error, QLIKE, and $R^2$ from MZ-regressions. Panel A reports errors for an average firm, Panel B reports error for an average cross-section, and Panel C reports the pooled error.}
\label{model_scoring_statistics}
\begin{tabular}{llllllll}
\hline
Forecast & RMSE &  MAE & QLIKE & $R^2_{\text{MZ Reg}}$ &  $\alpha_{\text{MZ Reg}}$ &  $\beta_{\text{MZ Reg}}$ \\
\hline
\hline
\multicolumn{7}{c}{\textbf{Panel A: Average Firm Error}} \\%
\hline
PCA + Lagged RVs &  0.001702 &  0.001202 &  3.186 & 0.4575 & -0.000016 &     0.2686 \\
Average Forecasts &  0.001051 &  0.000688  &  1.625 & 0.4866 & -0.000015 &     0.4000 \\
HAR &  0.000519 &  0.000200 &  0.2591 &  0.4692 &  0.000044 &     0.8763 \\
Rolling SD Squared &  0.000633 &  0.000306 &  0.6449 & 0.3229 & -0.000050 & 1.230 \\
LASSO &  0.000549 &  0.000217 &  0.3448 & 0.3906 & 0.000051 &     0.8763  \\
\hline
\hline
\multicolumn{7}{c}{\textbf{Panel B: Average Cross-Section of Firms Error}} \\ %
\hline
PCA + Lagged RVs &  0.002071 &  0.001392 &  3.437 & 0.2598 & 0.000112 &     0.2023\\
Average Forecasts &  0.001276 &  0.000810  &  1.835 & 0.3135 & 0.000015 &     0.3752 \\
HAR &  0.000728 &  0.000254 &  0.3309 & 0.3146 &  0.000081 &  0.8042 \\
Rolling SD Squared &  0.000841 &  0.000371 &  0.8299 & 0.0949 & 0.000212 & 0.6835  \\
LASSO &  0.000730 &  0.000275 &  0.4676 & 0.2804 & 0.000121 & 0.7942 \\
\hline
\hline
\multicolumn{7}{c}{\textbf{Panel C: Average Pooled Error }} \\
\hline
PCA + Lagged RVs &  0.002563 &  0.001192 &  3.179 &       0.2318 &      0.000007 & 0.2555 \\
Average Forecasts &  0.001843 &  0.000682 &  1.621 &      0.2378 &      0.000040 & 0.3519 \\
HAR &  0.001593 &  0.000198 &  0.2581 &     0.1223 &      0.000264 &     0.3418 \\
Rollling SD Squared &  0.001335 &  0.000304  & 0.6449 &      0.0768 &      0.000107 &     0.7175 \\
LASSO &  0.001348 &  0.000214 &  0.3374 &  0.1611 & 0.000184 &    0.5424 \\
\hline
\end{tabular}
\end{table}

%% file: tables/options_sample_summary_stats.tex
\begin{table}[htb]
\centering
\ra{1.3}
\caption[Options sample summary statistics.]{Panel A shows the average excess returns of straddles in our sample, which are, unconditionally, slightly negative. The computed sample moments are the time-series average for the moments of the cross-sectional distribution of the excess straddle returns. The mean gives the average return of an equal-weighted portfolio of all the straddles in our sample. Panel B gives the average cross-section of firms' summary statistics, and panel C gives the summary statistics of the average cross-section of firms' sorting variables, dependent upon the forecasting model.}
\label{options_summary_stats_variables}
\begin{tabular}{lllll}
\hline
Option Variables & Mean & SD & Skew & Kurtosis\\
\hline
\hline
\multicolumn{5}{c}{\textbf{Panel A: Unconditional Excess Returns}} \\
\hline
ATM Straddle  & -0.00006 &  0.8336 &  4.23 &  50.9 \\
\hline
\hline 
\multicolumn{5}{c}{\textbf{Panel B: Volatility Variables}} \\
\hline
ATM Straddle Implied Volatility & 0.3274 &  0.1164 &  1.59 & 5.64 \\
Realized Volatility & 0.0165 & 0.0071 & 2.44 & 12.9 \\
$\widehat{RV}$ - PCA + Lagged RVs & 0.0015 &  0.0015 &  4.25 & 29.4 \\
$\widehat{RV}$ - Average Forecasts & 0.0010 &  0.0010 &  4.21 & 28.3 \\
$\widehat{RV}$ - HAR & 0.0004 &  0.0005 &  6.80 &  78.4 \\
$\widehat{RV}$ - Rolling SD Squared & 0.0004 &  0.0003 &  2.50 & 7.12 \\
$\widehat{RV}$ - LASSO & 0.0004 &  0.0004 &  6.40 & 73.2 \\
\hline
\hline 
\multicolumn{5}{c}{\textbf{Panel C: Option Return Conditioning Variable - $\ln(RV/IV)$}} \\
\hline
VRP - PCA + Lagged RVs & 0.5082 &  0.2008 &  -0.28 &        3.62 \\
VRP - Average Forecasts & 0.1410 &  0.1750 &  -.0018 &        4.40 \\
VRP - HAR & -0.1664 &  0.1961 &  0.0373 & 4.91 \\
VRP - Rolling SD Squared & -0.0730 &  0.1777 &  -0.2087 & 5.76 \\
VRP - LASSO & -0.1589 &  0.2179 &  -0.1653 &  5.18 \\
\hline
\end{tabular}
\end{table}

%% file: tables/portfolio_stats_sharpe_mean.tex
\begin{table}[htb]
\centering
\ra{1.3}
\caption[Sorted portfolio performance metrics.]{We consider the average excess return to equally-weighted portfolios formed on the spread between a forecast of realized volatility and implied volatility. Panel A gives risk adjusted expected excess returns, and panel B gives the raw or absolute expected excess returns to the sorted straddle portfolios. These numbers are all unannualized at the daily frequency.}
\label{portfolio_stats_sharpe}
\begin{tabular}{llllllll}
\hline
Forecast & 1 & 2 & 3 & 4 & 5 &  5 minus 1 \\
\hline
\hline
\multicolumn{7}{c}{\textbf{Panel A: Sharpe Ratio of Straddle Portfolios}} \\
\hline
PCA + Lagged RVs & -0.3249 & -0.1299 & -0.0251 &  0.0940 &  0.2665 & 0.8044 \\
Average Forecasts & -0.2971 & -0.1192 & -0.0242 &  0.0776 &  0.2613 & 0.7876 \\
HAR & -0.1967 & -0.0876 & -0.0221 &  0.0567 &  0.2043 & 0.6317 \\
Rolling SD Squared & -0.2171 & -0.0799 & -0.0148 &  0.0572 &  0.1902 & 0.5769 \\
LASSO & -0.1706 & -0.0675 & -0.0138 &  0.0469 &  0.1701 & 0.5744 \\
\hline
\hline
\multicolumn{7}{c}{\textbf{Panel B: Average Straddle Portfolio Excess Returns}} \\
\hline
PCA + Lagged RVs & -0.0097 & -0.0043 & -0.0009 &  0.0035 &  0.0109 & 0.0206 \\
Average Forecasts & -0.0089 & -0.0041 & -0.0009 &  0.0029 &  0.0105 & 0.0195 \\
HAR & -0.0063 & -0.0031 & -0.0008 &  0.0021 &  0.0076 & 0.0139 \\
Rolling SD Squared & -0.0068 & -0.0028 & -0.0005 &  0.0021 &  0.0075 & 0.0143 \\
LASSO & -0.0056 & -0.0023 & -0.0005 &  0.0017 &  0.0063 & 0.0118 \\
\hline
\end{tabular}
\end{table}

%% file: tables/top_performing_portfolio_summary_stats.tex
\begin{table}[htb]
\centering
\ra{1.3}
\caption[Summary statistics of straddle portfolios for different forecasts.]{Summary statistics of the time-series of straddle portfolios formed from different forecasts.}
\label{summary_stats_straddle_returns}
\begin{tabular}{lllll}
\hline
Forecast &   Mean &    SD &  Skew &  Kurtosis \\
\hline
\hline
\multicolumn{5}{c}{\textbf{High-Low Sorted Straddle Portfolio Returns}} \\
\hline
PCA + Lagged RVs   &  0.0206 &  0.0256 &  2.11 &  18.64 \\
Average Forecasts  &  0.0195 &  0.0248 &  1.75 &  14.47 \\
HAR                &  0.0139 &  0.0221 &  0.603 &   8.04 \\
Rolling SD Squared &  0.0143 &  0.0248 &  0.966 &   7.35 \\
LASSO              &  0.0118 &  0.0206 &  0.501 &   6.33 \\
\hline
\end{tabular}
\end{table}